\documentclass[aps,prx,twocolumn,superscriptaddress,longbibliography]{revtex4-2}

\usepackage[a4paper,left=1.9cm,right=1.9cm,top=1.7cm,bottom=1.82cm]{geometry}

\usepackage{graphicx}
\usepackage{amsmath}
\usepackage{amssymb}
\usepackage[usenames,dvipsnames,svgnames,table]{xcolor}
\usepackage{hyperref} 
\hypersetup{colorlinks=true,linktoc=all,linkcolor=blue,breaklinks=true,citecolor=blue,urlcolor=blue}

\newcommand{\expval}[1]{\langle#1\rangle}
\newcommand{\order}{\mathcal{O}}

\newcommand{\vz}{\mathbf{\hat{z}}}

\renewcommand{\d}{\mathrm{d}}

\newcommand{\ket}[1]{\left|#1\right\rangle}
\newcommand{\bra}[1]{\left\langle#1\right|}

\newcommand{\kB}{k_{\mathrm{B}}}
\newcommand{\kBT}{\kB{}T}

\newcommand{\vS}{\mathbf{S}}
\newcommand{\Sabs}{S}
\newcommand{\SO}{S_{0}}
\newcommand{\Sx}{S_{x}}
\newcommand{\Sy}{S_{y}}
\newcommand{\Sz}{S_{z}}

\newcommand{\Sj}{S_{j}}
\newcommand{\Sk}{S_{k}}
\newcommand{\Sl}{S_{l}}
\newcommand{\Stheta}{S_{\theta}}

\newcommand{\sx}{s_x}

\newcommand{\sz}{s_z}

\newcommand{\sk}{s_k}
\newcommand{\stheta}{s_{\theta}}

\newcommand{\Sm}{S_{-}}
\newcommand{\Sp}{S_{+}}

\newcommand{\Bext}{B_{\mathrm{ext}}}

\newcommand{\vBext}{\mathbf{B_{\mathrm{ext}}}}
\newcommand{\wL}{\omega_{\mathrm{L}}}

\newcommand{\Xop}{X}
\newcommand{\Pop}{P}
\newcommand{\Xw}{\Xop_{\omega}}
\newcommand{\Pw}{\Pop_{\omega}}

\newcommand{\Xwn}{\Xop_{\omega_n}}
\newcommand{\Pwn}{\Pop_{\omega_n}}
\newcommand{\pw}{p_{\omega}}
\newcommand{\pwprime}{p_{\omega'}}
\newcommand{\qw}{q_{\omega}}
\newcommand{\xw}{x_{\omega}}

\newcommand{\Cw}{C_{\omega}} 
\newcommand{\Jw}{J_{\omega}} 
\newcommand{\Cwn}{C_{\omega_n}} 
 
\newcommand{\alp}{\alpha} 
\newcommand{\Q}{Q}
\newcommand{\Bop}{B}
\newcommand{\relstr}{\zeta}
\newcommand{\Jrc}{J_{\mathrm{RC}}}
\newcommand{\rcfreq}{\Omega_{\mathrm{RC}}}
\newcommand{\rcint}{\lambda_{\mathrm{RC}}}
\newcommand{\rcdis}{\gamma_{\mathrm{RC}}}

\newcommand{\HS}{H_{\mathrm{S}}}
\newcommand{\Hany}{H} 
\newcommand{\HR}{H_{\mathrm{R}}}
\newcommand{\Htot}{H_{\mathrm{tot}}} 

\newcommand{\Hint}{H_{\mathrm{int}}}
\newcommand{\Heff}{H_{\mathrm{eff}}}

\newcommand{\hR}{h_{\mathrm{R}}}
\newcommand{\hRshift}{\hR^{\mathrm{shift}}}
\newcommand{\HRCtot}{\Htot^{\mathrm{RC}}}
\newcommand{\HRC}{H_{\mathrm{RC}}}
\newcommand{\HRCint}{\Hint^{\mathrm{RC}}}
\newcommand{\HRes}{H_{\mathrm{res}}}
\newcommand{\HResint}{\Hint^{\mathrm{res}}}

\newcommand{\ttotMFGS}{\tau_{\mathrm{tot}}}
\newcommand{\tsysMFGS}{\tau_{\mathrm{MF}}}
\newcommand{\tauS}{\tau_{\mathrm{S}}}

\newcommand{\rhowkun}{\tilde{\rho}^{(2)}}
\newcommand{\rhowk}{\tau_{\mathrm{MF}}^{(2)}}
\newcommand{\rhoun}{\tilde{\rho}}

\newcommand{\Z}{Z}
\newcommand{\ZO}{\Z_0}
\newcommand{\ZR}{\Z_{\mathrm{R}}}
\newcommand{\ZS}{\Z_{\mathrm{S}}}
\newcommand{\ZSR}{\Z_{\mathrm{SR}}}
\newcommand{\Ztot}{\Z_{\mathrm{tot}}}
\newcommand{\ZtotMFGS}{\Z_{\mathrm{tot}}}
\newcommand{\ZRqu}{\ZR^{\quant}}
\newcommand{\ZRcl}{\ZR^{\class}}
\newcommand{\ZSRqu}{\ZSR^{\quant}}
\newcommand{\ZSRcl}{\ZSR^{\class}}

\newcommand{\Ztotqu}{\Ztot^{\quant}}
\newcommand{\ZScl}{\ZS^{\class}}
\newcommand{\ZSqu}{\ZS^{\quant}}
\newcommand{\ZMF}{\tilde{Z}}
\newcommand{\ZMFcl}{\ZMF^{\class}}
\newcommand{\ZMFqu}{\ZMF^{\quant}}
\newcommand{\ZSMF}{\tilde{\Z}_{\mathrm{S}}}
\newcommand{\ZSMFcl}{\ZSMF^{\class}}
\newcommand{\ZSMFqu}{\ZSMF^{\quant}}
\newcommand{\ZSMFwk}{\ZSMF^{(2)}}

\newcommand{\ZCMFUS}{\ZMFcl_{\mathrm{S,US}}}
\newcommand{\ZQMFUS}{\ZMFqu_{\mathrm{S,US}}}

\DeclareMathOperator{\tr}{tr}
\newcommand{\trR}{\tr_{\mathrm{R}}}
\newcommand{\trS}{\tr_{\mathrm{S}}}
\newcommand{\trSR}{\tr_{\mathrm{SR}}}
\newcommand{\trRcl}{\trR^{\class}}
\newcommand{\trRqu}{\trR^{\quant}}
\newcommand{\trScl}{\trS^{\class}}

\newcommand{\trSRqu}{\trSR^{\quant}}
\newcommand{\trqu}{\tr^\quant}

\newcommand{\IntW}{\int_{0}^{\infty}\hspace{-0.6em}\mathrm{d}\omega\,}

\newcommand{\IntPhi}{\int_{0}^{2\pi}\hspace{-0.6em}\mathrm{d}\varphi\,}
\newcommand{\IntTheta}{\int_{0}^{\pi}\hspace{-0.6em}\mathrm{d}\vartheta\,}

\newcommand{\dlg}{dashed light green}
\newcommand{\dlb}{dashed turquoise}
\newcommand{\dg}{dashed grey}
\newcommand{\sdb}{solid dark blue}
\newcommand{\dbt}{dark blue triangles}

\newcommand{\class}{\mathrm{cl}}
\newcommand{\quant}{\mathrm{qu}}
\newcommand{\MF}{MF}
\newcommand{\MFGS}{MF}
\newcommand{\CSS}{CSS}

\newcommand{\QMFWK}{\mbox{QMF}WK}
\newcommand{\CMF}{CMF}
\newcommand{\QMF}{QMF}
\newcommand{\QSS}{QSS}

\newcommand{\QMFUS}{\mbox{QMF}US}

\newcommand{\QG}{QG}
\newcommand{\UW}{UW}
\newcommand{\WK}{WK}
\newcommand{\IM}{IM}
\newcommand{\US}{US}
\newcommand{\supp}{\blue{Appendix}}
\newcommand{\supptracing}{A}

\newcommand{\suppqutocl}{D}

\newcommand{\suppus}{G}

\newcommand{\blue}[1]{\color{black}{#1}}

\newcommand{\exeter}{Department of Physics and Astronomy, University of Exeter,
                     Stocker Road, Exeter EX4 4QL, UK}
\newcommand{\oxford}{Department of Materials, University of Oxford, Parks Road,
                     Oxford OX1 3PH, United Kingdom}
\newcommand{\glasgow}{School of Physics and Astronomy, University of Glasgow,
                      Glasgow, G12 8QQ, UK}
\newcommand{\macquire}{Department of Physics and Astronomy, Macquarie University,
                       2109 NSW, Australia}
\newcommand{\potsdam}{Institut f{\"u}r Physik und Astronomie, University of Potsdam,
                      14476 Potsdam, Germany}

\begin{document}


\title{Quantum--classical correspondence in spin--boson equilibrium states
  \texorpdfstring{\\}{} at arbitrary coupling}


\author{F. Cerisola}
\email{federico.cerisola@eng.ox.ac.uk}
\affiliation{\exeter}
\affiliation{\oxford}

\author{M. Berritta}
\affiliation{\exeter}

\author{S. Scali}
\affiliation{\exeter}

\author{S.A.R. Horsley}
\affiliation{\exeter}

\author{J.D. Cresser}
\affiliation{\exeter}
\affiliation{\glasgow}
\affiliation{\macquire}

\author{J. Anders}
\email{janet@qipc.org}
\affiliation{\exeter}
\affiliation{\potsdam}


\date{\today}


\begin{abstract}

The equilibrium properties of nanoscale systems can deviate significantly from standard thermodynamics due to their coupling to an
environment.
For the generalised $\theta$-angled spin-boson model, \blue{we first derive a compact and general} form of the classical equilibrium state \blue{including environmental  corrections to all orders.}
Secondly, for the quantum spin-boson model we prove, by carefully taking a large spin limit, that Bohr's quantum-classical correspondence persists at all coupling strengths.
This correspondence gives insight into the conditions for a \blue{coupled} quantum spin to be well-approximated by a \blue{coupled classical spin-vector.}
\blue{Thirdly, we demonstrate that previously identified environment-induced `coherences' in the equilibrium state of weakly coupled quantum spins, do \textit{not} disappear in the classical case.}
\blue{Finally, we provide the first classification of the coupling parameter regimes for the spin-boson model, from weak to ultrastrong, both for the quantum case and the classical setting.}
Our results shed light on the interplay of quantum and mean force corrections in
equilibrium states of the spin-boson model, and will help draw the quantum to
classical boundary in a range of fields, such as magnetism and exciton dynamics.
\end{abstract}


\maketitle


Bohr's correspondence principle~\cite{Bohr1920} played an essential role in the
early development of quantum mechanics.
Since then, a variety of interpretations and applications of the correspondence
principle have been
explored~\cite{millard1971,lieb1973,liboff1975,liboff1984,kryvohuz2005,graefe2010,jarzynski2015,chen2021}.
One form asks if the statistical properties of a quantum system approach those
of its classical counterpart in the limit of large quantum
numbers~\cite{liboff1975,liboff1984}.
This question was answered affirmatively by Millard and Leff, and Lieb for a
quantum spin system~\cite{millard1971,lieb1973}. They proved that the system's
thermodynamic partition function $\ZSqu$ associated with the Gibbs state,
converges to the corresponding classical partition function $\ZScl$, in the
limit of large spins.
Such correspondence gives insight into the conditions for a quantum
thermodynamic system to be well-approximated by its classical
counterpart~\cite{jarzynski2015,chen2021}.
While $\ZSqu$ is computationally tough to evaluate for many systems, $\ZScl$
offers tractable expressions with which thermodynamic properties, such as free
energies, susceptibilities and correlation functions, can readily be
computed~\cite{millard1971,lieb1973}.
Similarly, many dynamical approaches solve a classical problem rather than the
much harder quantum problem. For example, sophisticated atomistic simulations of
the magnetisation dynamics in magnetic
materials~\cite{evans2014,vampire,barker2019,strungaru2021,barker2021} solve the
evolution of millions of interacting classical spins. A corresponding quantum
simulation~\cite{Vorndamme2021} would require no less than a full--blown quantum
computer as its hardware.

Meanwhile, in the field of quantum thermodynamics, extensive progress has
recently been made in constructing a comprehensive framework of “strong coupling
thermodynamics” for
classical~\cite{jarzynski2004,seifert2016,jarzynski2017,strasberg2017,miller2017,aurell2018}
and
quantum~\cite{mori2008,campisi2009b,hilt2011,fleming2011,thingna2012,subasi2012,philbin2016,miller2018,strasberg2019,trushechkin2021a,trushechkin2021b}
systems. This framework extends standard thermodynamic relations to systems
whose coupling to a thermal environment can not be neglected.
The equilibrium state is then no longer the quantum or classical Gibbs state,
but must be replaced with the environment-corrected mean force (Gibbs)
state~\cite{cresser2021a,trushechkin2021a,trushechkin2021b}.
These modifications bring into question the validity of the correspondence
principle when the environment-coupling is no longer negligible. Mathematically,
the challenge is that in addition to tracing over the system, one must also
evaluate the trace over the environment.

Strong coupling contributions are present for both classical and quantum
systems. However, a quantitative characterisation of the difference between
these two predictions, in various coupling regimes, is missing.
For example, apart from temperature, what are the parameters controlling the
deviations between the quantum and classical spin expectation values? %
Do coherences, found to persist in the mean force equilibrium state of a quantum
system~\cite{purkayastha2020}, decohere when taking the classical limit? 
\blue{How strong does the environmental coupling need to be for the spin-boson model to be well-described by weak or ultrastrong coupling approximations?}
In this paper, we answer these questions for the particular case of a spin $\SO$ coupled to a one-dimensional bosonic environment such that both dephasing and
detuning can occur ($\theta$-angled spin-boson model).


\smallskip

\noindent\textbf{Setting.}
This generalised version of the spin-boson model~\cite{ferialdi2017a,chiu2021}
describes a vast range of physical contexts, including excitation energy
transfer processes in molecular aggregates described by the Frenkel exciton
Hamiltonian~\cite{trushechkin2019,yang2002,kolli2012,moix2012,huelga2013,seibt2017,gelzinis2020},
the electronic occupation of a double quantum dot whose electronic dipole moment
couples to the substrate phonons in a semi-conductor~\cite{purkayastha2020}, 
an electronic, nuclear or effective spin exposed to a magnetic field and
interacting with an (anisotropic) phononic, electronic or magnonic
environment~\cite{anders2020,unikandanunni2022,neeraj2021,neeraj2022,stupakiewicz2021}, and a plethora of other aspects of quantum dots,
ultracold atomic impurities, and superconducting
circuits~\cite{nazir2009,recati2005,magazzu2018,popovic2021}.
In all these contexts, an effective ``spin'' $\vS$ interacts with an
environment, where $\vS$ is a vector of operators (with units of angular
momentum) whose components fulfil the angular momentum commutation relations
$[\Sj,\Sk] = i\hbar \sum_l \epsilon_{jkl}\Sl$ with $j,k,l=x,y,z$. We will
consider spins of any length $\SO$, i.e. $\vS^2 = \SO(\SO+\hbar)\boldsymbol{1}$.
The system Hamiltonian is
\begin{equation} \label{eq:HS}
  \HS = - \wL\Sz,
\end{equation}
where the system energy level spacing is $\hbar\wL > 0$ and the energy axis is
in the $-z$-direction without loss of generality. For a double quantum dot, the
frequency $\wL$ is determined by the energetic detuning and the tunneling
between the dots~\cite{purkayastha2020}.
For an electron spin with $\SO = \hbar/2$, the energy gap is set by a (negative)
gyromagnetic ratio $\gamma$ and an external magnetic field $\vBext = -\Bext\vz$,
such that $\wL = \gamma\Bext$ is the Larmor frequency.

\begin{figure}[tb]
  \centering
  \includegraphics[width=0.65\linewidth]{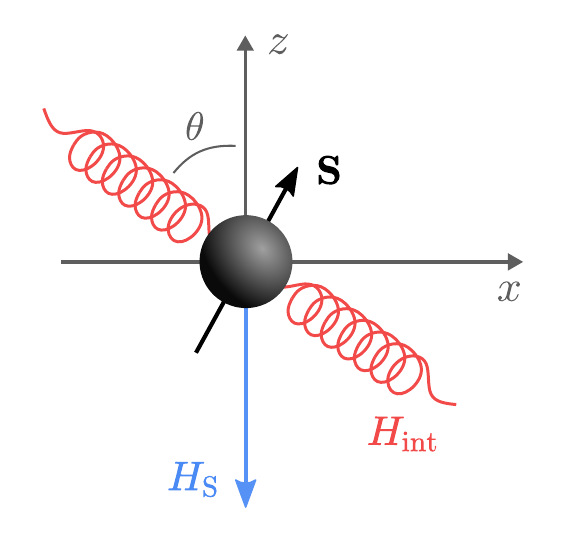}
  \caption{\sf \textbf{Illustration of bare and interaction energy axes.} A spin
  operator (vector) $\vS$ with system Hamiltonian $\HS$ with energy axis in
  the $-z$-direction is coupled in $\theta$-direction to a \blue{harmonic} 
  environment via $\Hint$.
  \label{fig:model}}
\end{figure}

The spin system is in contact with a bosonic reservoir, which is responsible for
the dissipation and equilibration of the system. Typically, this environment
will consist of phononic modes or an electromagnetic
field~\cite{breuer2007,trushechkin2021b}. The bare Hamiltonian of the reservoir
is
\begin{equation} \label{eq:HR}
  \HR = \frac{1}{2} \IntW \left(\Pw^2 + \omega^2\Xw^2\right),
\end{equation}
where $\Xw$ and $\Pw$ are the position and momentum operators of the reservoir
mode at frequency $\omega$ which satisfy the canonical commutation relations
$[\Xw,\Pop_{\omega'}] = i\hbar \, \delta(\omega - \omega')$.
With the identifications made in~\eqref{eq:HS} and~\eqref{eq:HR}, the
system-reservoir Hamiltonian is
\begin{equation} \label{eq:Htot}
  \Htot = \HS + \HR + \Hint,
\end{equation}
which contains a system-reservoir coupling $\Hint$. Physically, the coupling can
often be approximated to be linear in the canonical reservoir
operators~\cite{trushechkin2021b}, and is then modelled
as~\cite{huttner1992,breuer2007,purkayastha2020}
\begin{equation} \label{eq:Hint}
  \Hint = \Stheta \IntW \Cw\Xw,
\end{equation}
where the coupling function $\Cw$ determines the interaction strength between
the system and each reservoir mode $\omega$. $\Cw$ is related to the reservoir
spectral density $\Jw$ via $\Jw = \Cw^2/(2\omega)$.
It is important to note that the coupling is to the spin (component) operator
$\Stheta = \Sz\cos\theta - \Sx\sin\theta$ which is at an angle $\theta$ with
respect to system's bare energy axis, see Fig.~\ref{fig:model}. For example,
for a double quantum dot~\cite{purkayastha2020}, the angle $\theta$ is
determined by the ratio of detuning and tunnelling parameters.

In what follows we will need an integrated form of the spectral density, namely 
\begin{equation} \label{eq:reorg}
  \Q = \IntW \frac{\Jw}{\omega}
     = \IntW \frac{\Cw^2}{2\omega^2}.
\end{equation}
This quantity is a measure of the strength of the system--environment coupling
and it is sometimes called ``reorganization
energy''~\cite{fruchtman2016,wu2010,ritschel2011,cresser2021a}.
The analytical results discussed below are valid for arbitrary coupling
functions $\Cw$ (or reorganisation energies $\Q$). 
\blue{The plots assume Lorentzians
$\Jw = (2\Gamma \,\Q/\pi) \, \omega_0^2\omega/((\omega_0^2 - \omega^2)^2 + \Gamma^2\omega^2)$,  where $\omega_0$ is the resonant frequency of the Lorentzian~\cite{anders2020} and $\Gamma$ the peak width.}


\smallskip

We will model $\Htot$ (Eq.~\eqref{eq:Htot}) either fully quantum mechanically as
detailed above, or fully classically. To obtain the classical case, the spin
$\vS$ operator will be replaced by a three-dimensional vector of length $\SO$,
and the reservoir operators $\Xw$ and $\Pw$ will be replaced by classical
phase--space coordinates. Below, we evaluate the spin's so-called mean force
(Gibbs) states, \CMF{} and \QMF{}, for the classical and quantum case,
respectively.
The mean force approach postulates~\cite{trushechkin2021b} that the equilibrium
state of a system in contact with a reservoir at temperature $T$ is the mean
force (\MF) state, defined as
\begin{equation} \label{eq:MFGS}
  \tsysMFGS := \trR[\ttotMFGS]
             = \trR\left[\frac{e^{-\beta \Htot}}{\ZtotMFGS}\right].
\end{equation}
That is, $\tsysMFGS$ is the reduced system state of the global Gibbs state
$\ttotMFGS$, where $\beta = 1/\kBT$ is the inverse temperature with $\kB$ the
Boltzmann constant, and $\ZtotMFGS$ is the global partition function. Quantum
mechanically, $\trR$ stands for the operator trace over the reservoir space
while classically, ``tracing'' is done by integrating over the reservoir degrees
of freedom. Further detail on classical and quantum tracing for the spin and the
reservoir, respectively, is given in \supp~\supptracing.

\begin{figure*}[t]
  \centering
  \includegraphics[width=0.95\linewidth]{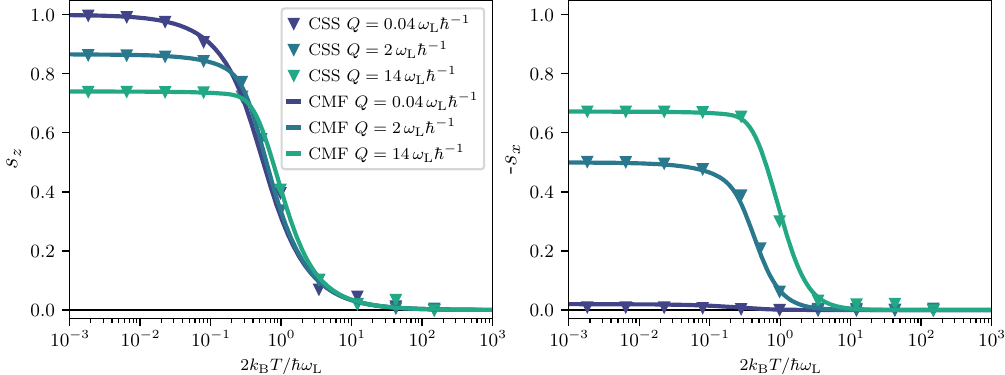}
  \caption{ \sf \blue{\textbf{Classical mean force and steady--state spin expectation values.}}
  Normalised expectation values of the classical spin components $\sz$
  \textbf{(left)} and $\sx$ \textbf{(right)} as a function of temperature.
  These are obtained with: (\CSS) the long time average of the dynamical
  evolution of the spin, $\sk = \expval{\Sk}_{\mathrm{\CSS}}/\SO$; and (\CMF)
  the classical \MFGS{} state (Eq.~\eqref{eq:taumf}), $\sk =
  \expval{\Sk}_{\mathrm{\MF}}/\SO$. These are shown for three different 
  coupling strengths
  $\Q = 0.04\,\wL\hbar^{-1}, 2\,\wL\hbar^{-1}, 14\,\wL\hbar^{-1}$,
  that range from the weak to the strong coupling regimes. In all three cases, we
  see that the \MFGS{} predictions are fully consistent with the results of the
  dynamics. All plots are for Lorentzian coupling with $\omega_0 = 7\wL$,
  $\Gamma = 5\wL$, and coupling angle $\theta = 45^\circ$.
  The temperature scale shown corresponds to a
  spin $\SO=\hbar/2$. 
  \label{fig:css}}
\end{figure*}

While the formal definition of $\tsysMFGS$ is deceptively simple, carrying out
the trace over the reservoir -- to obtain a quantum expression of $\tsysMFGS$ in
terms of system operators alone -- is notoriously difficult. Often, expansions
for weak coupling are made~\cite{mori2008,purkayastha2020}. For a general
quantum system (i.e. not necessarily a spin), an expression of $\tsysMFGS$ has
recently been derived in this limit~\cite{cresser2021a}. Furthermore, recent
progress has been made on expressions of the quantum $\tsysMFGS$ in the limit of
ultrastrong coupling~\cite{cresser2021a}, and for large but finite
coupling~\cite{trushechkin2021a,trushechkin2021b,latune2021}. Moreover, high
temperature expansions have been derived that are also valid at intermediate
coupling strengths~\cite{gelzinis2020}. However, the low and
intermediate temperature form of the quantum $\tsysMFGS$ for intermediate
coupling is not known, neither in general nor for the $\theta$--angled spin
boson model~\cite{Nick2022}.

\medskip

\noindent\textbf{Classical \MF{} state at arbitrary coupling.}
In contrast, here we \blue{establish} that the analogous problem of a \textit{classical}
spin vector of arbitrary length $\SO$, coupled to a \blue{harmonic} reservoir via
Eq.~\eqref{eq:Hint}, is tractable for arbitrary coupling function $\Cw$ and
arbitrary temperature.
By carrying out the (classical) partial trace over the reservoir, i.e.
$\trRcl[\ttotMFGS]$, we uncover a rather
compact expression for the spin's \CMF{} state $\tsysMFGS$ and the \CMF{}
partition function $\ZSMFcl$:
\begin{align} \label{eq:taumf}
  \tsysMFGS &= \frac{e^{-\beta(\HS - \Q\Stheta^2)}}{\ZSMFcl} \\
    \text{with} \quad 
  \ZSMFcl &= \trScl[e^{-\beta(\HS - \Q\Stheta^2)}].
    \nonumber
\end{align}
The state $\tsysMFGS$ clearly differs from the standard Gibbs state by the
presence of the reorganisation energy term $-\Q\Stheta^2$.
The quadratic dependence on $\Stheta$ changes the character of the distribution,
from a standard exponential to an exponential with a positive quadratic term,
altering significantly the state whenever the system--reservoir coupling is
non-negligible.

\begin{figure*}[t]
  \centering
  \includegraphics[width=0.95\linewidth]{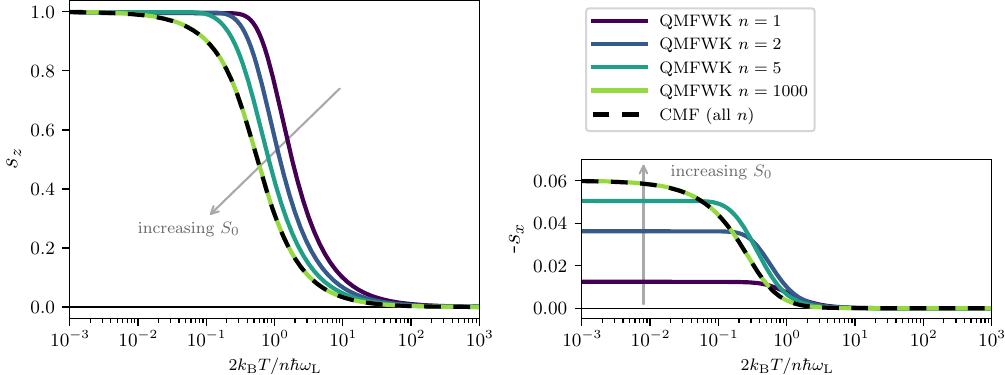}
  \caption{\sf \textbf{Classical and quantum mean force spin components.} 
  Normalised expectation values of the spin components $\sz$ \textbf{(left)} and
  $\sx$ \textbf{(right)} obtained with: (\QMFWK) the quantum \MF{} partition
  function $\ZSMFqu$ in the weak coupling limit for a spin of length $\SO =
  n\hbar/2$ ($n=1,2,5,100$);
  (\CMF) the classical \MF{} partition function $\ZSMFcl$ given
  in~\eqref{eq:taumf}. As the length $\SO$ of the quantum spin is increased, the
  quantum mean force prediction \QMFWK{} converges to that corresponding to the
  \CMF{} state.
  Non-zero $\sx$ \textbf{(right)} indicate ``coherences'' with respect to the
  system's bare energy axis ($z$). These arise entirely due to the
  spin-reservoir interaction. Such coherences have been discussed for the
  quantum case~\cite{purkayastha2020}. Here we find that they also arise in the
  classical \CMF{} and, comparing like with like for the same spin length
  $\SO=\hbar/2$, the classical ``coherences'' are \textit{larger} than those of
  the quantum spin.
  All plots are for a weak coupling strength, $\alp = 0.06$, and $\theta=\pi/4$.
  \label{fig:eqmag_vs_T}}
\end{figure*}

Throughout this article, we will consider that the \MF{} state is the
equilibrium state reached by a system in contact with a thermal bath.
While this is widely thought to be the case, some open questions remain about
formal proofs showing the convergence of the dynamics towards the steady state
predicted by the \MF{} state~\cite{trushechkin2021b,fleming2011,thingna2012,subasi2012,purkayastha2020,jaksic1996,merkli2001,bach2000,merkli2021,merkli2022a,merkli2022b}. 
For example, for quantum
systems, this convergence has only been proven in the weak~\cite{mori2008} and
ultrastrong limits~\cite{trushechkin2021a}, while for intermediate coupling
strengths there is numerical evidence for the validity of the \MF{}
state~\cite{chiu2021}.
Here, we numerically verify the convergence of the dynamics towards the \MF{}
state for the case of the classical spin at arbitrary coupling strength. This is
possible thanks to the numerical method proposed in~\cite{anders2020}.
Fig.~\ref{fig:css} shows the long time average of the spin components
once the dynamics has reached steady state (\CSS, triangles), together with the
expectation values predicted by the static \MF{} state (\CMF, solid lines), for
a wide range of coupling strengths going from weak to strong
coupling~\footnote{See later discussions where the different regimes of coupling
strength are thoroughly characterised.}.
We find that both predictions are in excellent agreement, providing strong
evidence for the convergence of the dynamics towards the \MFGS.
\blue{The compact expression \eqref{eq:taumf} for the CMF state, as well as the numerical verification that the dynamical steady state matches it, are the first result of this paper.}

\medskip


\noindent\textbf{Quantum--classical correspondence.}
%
%
We now demonstrate that the quantum partition function $\tilde{Z}_\text{S}^\text{qu}$,
which includes arbitrarily large mean force corrections, converges to the classical one, 
$\tilde{Z}_\text{S}^\text{cl}$ in Eq.~\eqref{eq:taumf}.

A well-known classical limit of a quantum spin is to increase the quantum spin's length, $\SO \to \infty$. This is because, when $\SO$ increases, the quantised angular momentum
level spacing relative to $\SO$ decreases, approaching a continuum of states that can be described in terms of a classical vector~\cite{Bohr1920}.
Taking the large spin limit for a spin-$\SO$ system can be achieved following an approach used by Fisher when treating an uncoupled spin with Hamiltonian $\HS$~\cite{fisher1964}. This involves introducing a rescaling of the spin operators via $s_j = S_j/\SO$ so that the commutation rule
becomes $[s_j,s_k] = i\hbar \, \epsilon_{jkl} \, s_l/\SO$. Hence, in the limit of $\SO\to\infty$, the scaled operators will commute, so in that regard they can be considered as classical quantities~\cite{fisher1964}. 
Millard \& Leff ~\cite{millard1971}  take this further and prove, for any spin Hamiltonian $\Hany$ in the spin Hilbert space ${\cal H}_S$, the identity 
\begin{align}
  \label{eq:sm:31}
  & \lim_{\SO\to\infty} \frac{\hbar}{2\SO+\hbar}
    \, \trS^\quant \left[e^{-\beta \Hany}\right] \nonumber \\
  = &\lim_{\SO\to\infty}\frac{1}{4\pi} \int_{0}^{2\pi}d\varphi\int_{0}^{\pi}d\vartheta\sin \vartheta \,
    e^{-\beta \Hany (\SO,\vartheta,\varphi)}, 
\end{align}
provided the limit on the right hand side exists 
\footnote{The factor of $\hbar/(2\SO+\hbar)$ guarantees
that the sides of \eqref{eq:sm:31} are equal for $\beta = 0$.
For a fixed value of $\SO$,
this pre-factor is un-important as it immediately cancels in any
calculation of expectation values, i.e. for a quantum system, the expressions  $\frac{\hbar}{2\SO+\hbar}\ZSqu(\beta,\SO)$ and
$\ZSqu(\beta,\SO)$ give the same expectation values.}.
Here $\Hany (\SO,\vartheta,\varphi)$ is the classical spin-$\SO$ Hamiltonian, where the spin-vector $\vS$ is parametrised by two angles, $\varphi$ and $\vartheta$, such that $s_x = \SO\sin\vartheta\cos\varphi,
s_y = \SO\sin\vartheta\sin\varphi$ and $s_z = \SO\cos\vartheta$.
Eq.~\eqref{eq:sm:31} was further confirmed by Lieb who provides a rigorous argument based on the  properties of spin-coherent states~\cite{lieb1973}.

Note, though, if one simply takes the $\SO \to \infty$ limit in \eqref{eq:sm:31}, with $\Hany$ being the system Hamiltonian $H_S \propto \SO$, that would have the same effect as sending $\beta \to \infty$; namely, all population will go to the ground state. Instead, to maintain a non-trivial temperature dependence after taking the $\SO$-limit requires a further rescaling step.
One approach involves a rescaling of the physical parameters of the Hamiltonian $\Hany$, as followed, e.g., by Fisher~\cite{fisher1964}.
A second approach is to rescale the inverse temperature via
$\beta\SO = \beta'$, and take the limit $\SO \to \infty$ with
$\beta'$ held fixed. This is the limit we will take here.
The effect of this constrained limit can readily be seen for the thermal states of the uncoupled classical or quantum spin. The classical partition function $\ZScl(\beta \SO) = {\sinh(\beta \SO\wL)/\beta \SO\wL}$ is left invariant because $\beta$ and $\SO$ always appear together in $\ZScl$. 
In contrast, the quantum partition function $\ZSqu(\beta,\SO)=\sinh(\beta(\SO+\hbar/2) \wL) / \sinh(\beta\hbar\wL/2)$ is altered in the constrained limit, since $\ZSqu$ separately depends on $\beta$ and $\SO$.
Eq.~\eqref{eq:sm:31} then expresses the convergence of the partition functions~\cite{fisher1964,millard1971,lieb1973}, i.e. $\frac{\hbar}{2\SO+\hbar} \, \ZSqu(\beta,\SO) \to \ZScl (\beta \SO)$.

\smallskip 

We now take a step further and extend this result to the case of a spin {\it coupled to a reservoir}. 
The first step is to consider that the relevant Hilbert space is now the tensor product space of spin and reservoir degrees of freedom, ${\cal H}_S \otimes  {\cal H}_B$.  It was argued by Lieb~\cite{lieb1973}  that \eqref{eq:sm:31} remains valid in this case, i.e. even when $\Hany \in {\cal H}_S \otimes {\cal H}_B$.
This means we can replace $\Hany$  in  \eqref{eq:sm:31} by our $\Htot$. But note that the trace is still only over the system Hilbert space ${\cal H}_S$. 
Thus, formally one obtains an operator valued identity for operators on ${\cal H}_B$.
The second step is then to evaluate the trace over the reservoir degrees of freedom.
To do so, we start by writing the total unnormalised Gibbs state as
\begin{align}
  &e^{-\beta \Htot} = \\ \nonumber
  &\exp\left[-\beta'\left(-\wL\sz + \frac{\HR}{\SO} + \stheta \int_{0}^{\infty} \d \omega\, \Cw \Xw  \right)\right].
\end{align}
with the rescaled inverse temperature $\beta' = \beta\SO$. Since $\beta'$ is constant as the limit $\SO\to\infty$ is taken, doing so rescales the spin operators,
as required. But it also rescales $\HR$ to $\hR = \HR/\SO$, which can be expressed in terms of rescaled reservoir operators, $\pw$ and $\xw$ 
where $\pw  = \Pw/\sqrt{\SO}$ and $\xw  = \Xw/\sqrt{\SO}$.
The commutation relations are then $\left[\xw ,\pwprime\right] = i\hbar \, \delta(\omega-\omega')/\SO$,
so in the limit of $\SO\to\infty$, these two operators commute~\cite{Wang1973}.
Thus, the classical limit of the spin induces a  limit for the reservoir.
I.e., the quantum nature of the reservoir is inevitably stripped away, so that the result eventually obtained is that of a classical spin coupled to a classical reservoir.

Written in terms of these rescaled reservoir operators, one now has
\begin{align} \label{eq:rescaledH}
  &e^{-\beta\Htot} = \\ \nonumber
  &\exp\left[-\beta'\left(-\wL\sz + h_R + \stheta \int_{0}^{\infty}\mathrm{d}\omega\, \Cw \sqrt{\SO} \, \xw \right)\right].
\end{align}
If one were to naively take the $\SO$-limit, then the interaction term dominates and the dependence on the bare  system energy $-\wL\sz$ drops out.     
To maintain a non-trivial dependence on both, bare and interaction energies, one needs to make an assumption on the scaling of the coupling function $\Cw$ with spin-length $\SO$. 
We choose to keep the relative energy scales of the bare and interaction Hamiltonians the same throughout the $\SO$ limit. Eq.~\eqref{eq:rescaledH} shows that this requires a scaling of  $\Cw \propto 1/\sqrt{\SO}$. 
This implies a reorgansiation energy  \eqref{eq:reorg} scaling of 
\begin{equation}
    Q = \alpha\frac{\wL}{\SO},
    \label{eq:scalingchoice}
\end{equation}
where $\alpha$ is a unit-free constant independent of $\SO$ and $\beta$.
Inserted in the classical MF state~\eqref{eq:taumf} this shows that both, the system energy
$\HS$ as well as the correction that comes from the reservoir interaction, scale as $\SO$. 
The combined scaling of $Q$ with $\SO$ (Eq.~\eqref{eq:scalingchoice}), and the rescaling of the inverse temperature, $\beta\SO = \beta' = \mathrm{const}$, then leaves the CMF state~\eqref{eq:taumf} invariant under variation of $\SO$.

Crucially, given the same scaling, the QMF state defined by Eq.~\eqref{eq:MFGS} will not be invariant under variation of $\SO$. 
Returning to the unnormalised total Gibbs state \eqref{eq:rescaledH},
taking the quantum trace over the spin, and using Eq.~\eqref{eq:sm:31}, one obtains an identity that still contains the bath operators in contrast to the uncoupled spin. 
Finally taking the quantum trace over the reservoir on both sides,
one finds
\begin{eqnarray}
  && \lim_{\SO\to\infty} \frac{\hbar}{2\SO+\hbar} \frac{\ZSRqu(\beta,\SO, \alp)}{\ZRqu(\beta)} \nonumber \\ 
   = && \lim_{\SO\to\infty} \frac{\hbar}{2\SO+\hbar} \ZSMFqu(\beta,\SO, \alp) 
  = \ZSMFcl(\beta\SO, \alp). \quad
  \label{eq:sm:new35}
\end{eqnarray}
Here it was used that the fraction of the total quantum partition function divided by the bare quantum reservoir partition function is the quantum mean force partition function (see \supp~\suppqutocl)~\cite{campisi2009,strasberg2020}.
\blue{In contrast to the quantum-classical correspondence established by Millard \& Leff, and Lieb, for the standard Gibbs state partition functions, there now is a dependence on the spin-environment coupling strength $\alpha$. For the classical case, one has $\ZSMFcl(\beta\SO, \alp)=\trScl[e^{-\beta\SO(\HS/\SO - \alp \, \wL \Stheta^2/\SO^2)}]$.}

\blue{
While we derived Eq.~\eqref{eq:sm:new35}  assuming a constant ratio between bare and interaction energy, i.e.  $\Cw \propto 1/\sqrt{\SO}$, the quantum-classical correspondence also holds for other scalings. Indeed, when $\Cw \propto 1/\sqrt{\SO^p}$ with  $p > 1$ the bare energy will grow much more rapidly than the interaction term in the limit $\SO \to \infty$. This  immediately leads to the ultraweak coupling limit where the known quantum-classical correspondence \eqref{eq:sm:31} applies.
On the other hand, when  $\Cw \propto 1/\sqrt{\SO^p}$ with  $0 \le p < 1$, the interaction term will grow much more rapidly than
the bare energy. As we show in \supp~\suppus, in this ultrastrong limit~\cite{cresser2021a}, the quantum and classical mean force partition functions turn out to be identical. 
\blue{In this ultrastrong limit, the partition function loses all dependence on the coupling strength $\alpha$.}
%
Thus, while \eqref{eq:sm:new35} is valid for $p>1$ and $0\le p <1$, these scalings give a trivial correspondence, \blue{independent of $\alpha$}. Only for the scaling~\eqref{eq:scalingchoice} is a dependence of the mean force partition functions on the coupling strength retained. 

The results presented here show the quantum-classical correspondence of the equilibrium states of the spin-boson model for the first time. The proof of this correspondence, valid at all coupling strengths, is the second result of the paper. 

\smallskip

We remark that, in the above proof, it was assumed that $\alp$ is independent of $\beta$. Physically this is not entirely accurate because the coupling $\Cw$ is usually a function of temperature~\cite{nemati2021}, albeit often a rather weak one. For the same limiting process to apply, a weak dependence on $\beta$ would need to be compensated by an equally weak additional dependence of $\Q$ on $\SO$.
}

\color{black}

To visually illustrate the quantum to classical convergence, we choose a weak
coupling strength, $\alp = 0.06$, for which an analytical form of the quantum
$\ZSMFqu$ is known~\cite{cresser2021a}.
Mean force spin component expectation values $\sk \equiv
\expval{\Sk}_{\mathrm{\MF}}/\SO$ for $k = x,z$ can then readily be computed from the partition functions $\ZSMFqu$ and $\ZSMFcl$, respectively.
Fig.~\ref{fig:eqmag_vs_T} shows $\sz$ and $\sx$ for various spin lengths, $\SO =
n\hbar/2$ with $n = 1,2,5,1000$ for the quantum case (\QMFWK, purple to green)
and the classical case (\CMF, dashed black).
Note, that the $x$-axis is a correspondingly rescaled temperature,
$2\kBT/n\hbar\wL$, a scaling under which the \CMF{} remains invariant. The
numerical results illustrate that the quantum $\sz$ and $\sx$ change with spin
length $\SO = n\hbar/2$, and indeed converge to the classical prediction in the
large spin limit, $n \to \infty$.

\medskip

\begin{figure}[t]
  \centering
  \includegraphics[width=0.95\linewidth]{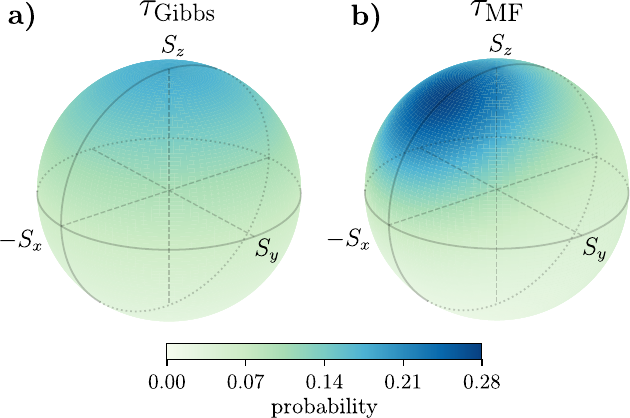}%
  \caption{\sf \blue{\textbf{Coherences and inhomogeneous probabilities.}
  Spin vector probability distributions (blue = high probability,
  white = low probability) as a function of three spin components on
  a sphere of radius $\SO$.
  \textbf{a)}~The classical Gibbs state 
  $\tau_\mathrm{Gibbs}$ is a homogeneous function of $\HS$, i.e. it is constant over
  the energy shells of $\HS$ which are fixed by the value of $\Sz$.
  \textbf{b)}~The classical mean force probability distribution $\tsysMFGS$ given 
  in~\eqref{eq:taumf}, peaks in a direction with positive components in $-\Sx$ and $\Sz$ directions.
  This makes $\tsysMFGS$ an inhomogeneous probability distribution over the energy
  shells of $\HS$.
  Parameters for the plots: $\theta=\pi/4$, $\kBT = \SO\wL$,
  and $\Q = \wL/\SO$.}}
  \label{fig:coherences}
\end{figure}

\begin{figure*}[t!]
  \centering
  \includegraphics[width=0.99\linewidth]{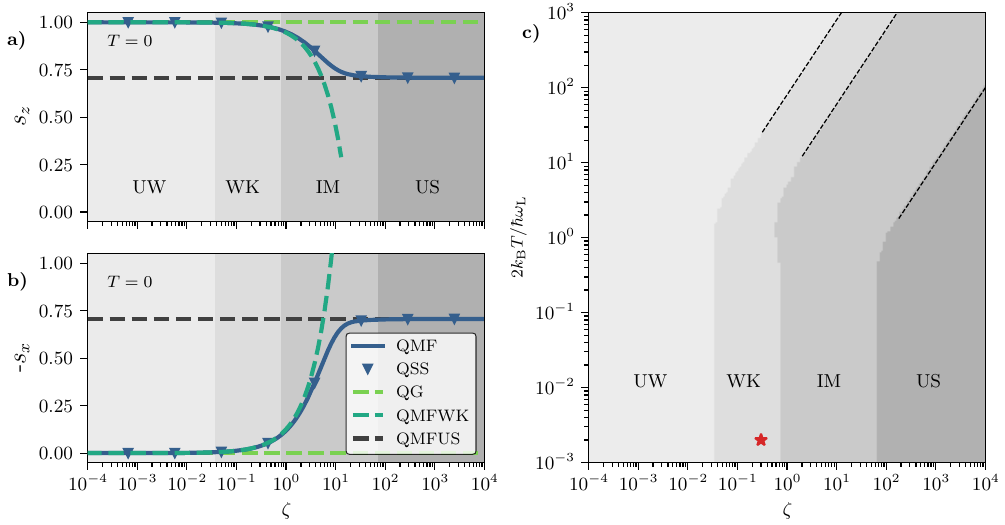}
  \caption{\sf \blue{\textbf{Quantum coupling regimes at $T=0$~K and $T>0$~K.}
  Panels \textbf{a)} and \textbf{b)}:
  Spin expectation values $\sz$ and $-\sx$ for the MF state \eqref{eq:MFGS} at $T=0$ for the total Hamiltonian \eqref{eq:Htot} with $\SO=\hbar/2$, $\theta=\pi/4$ and different coupling strengths as quantified by the dimensionless parameter $\relstr$, see~\eqref{eq:relstrdef}.
  We identify four coupling regimes for the numerically exact \QMF{} state  (\sdb): 
  \quad Ultraweak coupling (\UW), where the spin expectation values are consistent   with the Gibbs state (\QG, \dlg); 
 \quad Weak coupling (\WK), where the expectation values are well approximated by a second order expansion in $\relstr$ (\QMFWK, \dlb)~\cite{purkayastha2020};
\quad  Ultrastrong coupling (\US), where the  asymptotic limit of infinitely  strong coupling $\relstr \to \infty$ is valid (\QMFUS, \dg)~\cite{cresser2021a},  
\quad and Intermediate coupling (\IM) where the \QMF{} state is not approximated by any known analytical expression. 
   The  dynamical steady state of the quantum spin (\QSS, \dbt) is also computed using the reaction coordinate technique~\cite{iles-smith2014,iles-smith2016,nazir2018,correa2019}. Excellent agreement between the \QSS{} and the \QMF{} prediction is seen for all $\relstr$.
  Panel \textbf{c)}: Coupling regimes as a function of temperature $T$ and coupling strength $\relstr$. 
  With increasing temperature, the boundaries shift towards higher coupling $\relstr$. At large temperatures, all three boundaries follow a linear  relation $T \propto \relstr$ (dashed lines). }}
  \label{fig:regimes}
\end{figure*}

\begin{figure*}[t!]
  \centering
\includegraphics[width=0.99\linewidth]{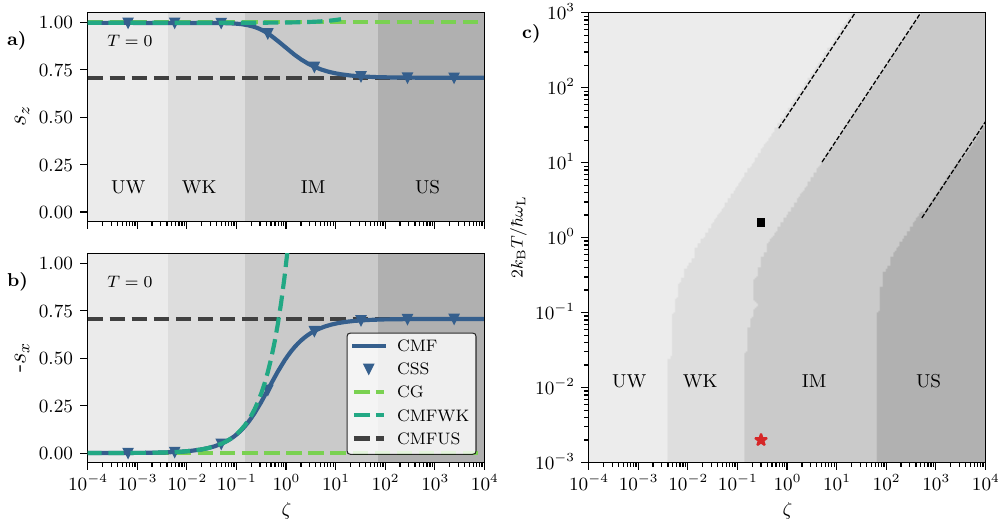}   \caption{\sf \blue{\textbf{Classical coupling regimes at $T=0$~K and $T>0$~K.} Same plot as Fig.~\ref{fig:regimes}, but here for the equilibrium state of a {\it classical} spin vector $\vS$ with Hamiltonian Eq.~\eqref{eq:Htot}.
A particular $(T,\relstr)$-pair (red star) is identified for which the classical spin falls in the intermediate regime. For the same parameter pair, the {\it quantum} spin falls in the weak coupling regime, see red star in Fig.~\ref{fig:regimes}c). 
Moving the classical red star upwards in temperature until it reaches a point (black square) in the weak coupling regime that is laterally distanced from the boundaries similar to the quantum star, Fig.~\ref{fig:regimes}c), gives an effective temperature shift of $\Delta T = 1.6 \cdot 2 \hbar \wL/\kB$.
This example evidences significant differences between the environmental impact on quantum and classical equilibrium states.
} 
}  \label{fig:clregimes}
  \end{figure*}


\noindent\textbf{\blue{Coherences.}}
\blue{
As seen in Fig.~\ref{fig:eqmag_vs_T} (right panel), the $\sx$ spin-component in the quantum mean force state (solid purple line for spin-1/2) is non-zero at low temperatures, despite the fact that the bare system energy scale is set along the $z$-direction, see~\eqref{eq:HS}. Such non-zero $\sx$ implies the presence of energetic ``coherences'' in the system's equilibrium state, as recently discussed in~\cite{guarnieri2018,purkayastha2020} for a quantum spin-1/2. 
Considering the quantum--classical correspondence discussed above, a natural
question is whether in the classical limit one can observe ``decoherence'', in
the sense of vanishing coherences. However, comparing like with like, we see in
Fig.~\ref{fig:eqmag_vs_T} that ``coherences'' are also present for
a classical spin with length $\SO = \hbar/2$ (dashed black). Indeed, maybe
surprisingly, classical ``coherences'' can be even larger in magnitude than
those of a quantum spin with corresponding length~$\SO$.
}

\blue{This observation reveals that the mechanism that gives rise to these
coherences is not an intrinsically quantum one.
Indeed, when we plot our CMF state~\eqref{eq:taumf} in Fig.~\ref{fig:coherences}b),  one can immediately see that the classical spin alignment in equilibrium tilts towards the $-x$ direction compared to the Gibbs state shown in Fig.~\ref{fig:coherences}a).
Such `inhomogenity' of a classical distribution has recently been identified by A. Smith, K. Sinha, C. Jarzynski~\cite{jarzynski2020} as the classical analogue to quantum coherences in the context of thermodynamic work extraction~\cite{kammerlander2016,hasegawa2010}. 
Here we uncover that the mechanism of producing such classical coherences can be due to the nature of the environmental coupling, which is asymmetric with respect to the bare Hamiltonian, see Fig.~\ref{fig:model}. 
\blue{This third finding, that coherences can be present even in classical equilibrium states, will have implications on a variety of fields, including quantum thermodynamics and quantum biology, which have so far interpreted a non-zero value of $\sx$ as a `quantum signature'.}
}

\medskip

\noindent\textbf{Coupling regimes.}
\blue{Finally, we now classify the interaction strength necessary for the spin-boson model to fall in various coupling regimes, from ultra-weak to ultra-strong. }
To quantify the relative strength of coupling we use the dimensionless parameter
\begin{equation} \label{eq:relstrdef}
  \relstr = \frac{\Q\SO}{\wL},
\end{equation}
which is the ratio of interaction and bare energy terms, see also Eq.~\eqref{eq:taumf}.
For the scaling choice~\eqref{eq:scalingchoice}, one has
$\relstr = \alp$.
It's important to note that temperature sets another scale in this problem --
higher temperatures will allow higher coupling values $\relstr$ to still fall
within the ``weak'' coupling regime~\cite{cresser2021a,latune2021b}. Thus, we
will first characterise various coupling regimes at $T = 0$~K, where the coupling
has the most significant effect on the system equilibrium state, and then
proceed to study finite temperatures.

\blue{
Fig.~\ref{fig:regimes}a) and~\ref{fig:regimes}b) show the spin components $\sz$ and $\sx$ in the quantum \MF{} state (\QMF, \sdb) at $T=0$~K. These expressions are evaluated numerically using the reaction coordinate
method~\cite{iles-smith2014,iles-smith2016,nazir2018,correa2019} for $\SO=\hbar/2$ and angle $\theta=\pi/4$.
Also shown are the spin components for the quantum Gibbs state (\QG, dashed green),
for the quantum \MF{} state in the weak coupling limit (\QMFWK, \dlb), and
for the quantum \MF{} state in the ultrastrong coupling limit (\QMFUS, dashed
grey)~\cite{cresser2021a}.

By comparing the analytical results (dashed lines) to the numerically exact result (solid line), and requiring the relative error to be at most $4\cdot10^{-3}$,  we can clearly identify four  regimes: \quad
For $\relstr < 4\cdot10^{-2}$, equilibrium is well-described by the quantum Gibbs state and this parameter regime can thus can be considered as ultraweak coupling (\UW)~\cite{trushechkin2021b}. 
For $4\cdot10^{-2} \le \relstr < 8\cdot10^{-1}$, equilibrium is well-described by the weak coupling state \QMFWK, which includes second order coupling corrections~\cite{cresser2021a}. Thus, this regime is identified as the weak coupling regime (\WK).
At the other extreme, for $7\cdot10^{1} \le \relstr$, the equilibrium state converges to the ultrastrong coupling state \QMFUS{} which was derived in~\cite{cresser2021a}.
Thus, this regime is identified as the ultrastrong coupling regime (\US).
Finally, for the parameter regime $8\cdot10^{-1} \le \relstr < 7\cdot10^{1}$ the exact \QMF{} shows variation with $\relstr$ that is not captured by neither weak nor ultrastrong coupling approximation. This is the intermediate coupling regime (\IM), which is highly relevant from an experimental point of view, but there are no known analytical expressions that approximate the numerically exact \QMF~\cite{Nick2022}. }

Beyond the zero temperature case, we 
compute $\sx$ and $\sz$ with the numerically exact \QMF{} state over a wide range of coupling strengths and temperatures, and
compare the results with those of the \UW, \WK, and \US{} approximations \blue{allowing an error of $4 \cdot 10^{-3}$, as above.}
\blue{Fig.~\ref{fig:regimes}c) shows how pairs of 
$\relstr$ and $T$ fall into various coupling regimes.} 
One can see that, at elevated temperatures, the coupling regime boundaries shift towards higher coupling $\relstr$. Thus at higher
temperatures, $2\kBT/\hbar\wL \gtrsim 10$, the \UW{} and \WK{} approximations
are valid at much higher coupling strengths $\relstr$ than at $T=0$. At higher
temperatures we also observe an emerging linear relation, $2\kBT/\hbar\wL
\propto \relstr$, for all three regime boundaries.
The temperature dependence of the border between the weak and intermediate coupling regime has previously been identified to be linear by C.~Latune~\cite{latune2021b}. 

\blue{
The quantum coupling regimes can now be compared to the corresponding regimes for a classical spin vector, shown in Fig.~\ref{fig:clregimes}a-c).
Perhaps surprisingly, we find that the classical regime boundary values for $\relstr$ differ significantly from those for the quantum spin, e.g., by a factor of 10 for
the \WK{} approximation.
This shift is exemplified by the red star, which indicates the same parameter pair $(T,\relstr)$ in both figures, Figs.~\ref{fig:regimes}c) and \ref{fig:clregimes}c). 
While the open quantum spin lies in the weak coupling regime, the classical one requires an intermediate coupling treatment.
We suspect this quantum-classical distinctness comes from the fact that, while for a classical spin at zero temperature there is no noise induced by the bath, in the quantum case noise is present even at $T=0$~K due to the bath's zero-point-fluctuations~\cite{anders2020}. One may qualitatively interpret this additional noise as an effective  temperature shift with respect to the classical case, by ca. $\Delta T = 1.6 \cdot  2 \hbar \wL/\kB$, as indicated by the black square in Fig.~\ref{fig:clregimes}c).}

\blue{To conclude, for any given coupling value $\relstr$ and temperature $T$, the two plots Fig.~\ref{fig:regimes}c) and Fig.~\ref{fig:clregimes}c) provide a tool to judge whether a ``weak coupling'' approximation is valid for the spin-boson model or not. We emphasise that, interestingly, the answer depends on whether one considers a  quantum or a classical spin. }

\medskip


\noindent\textbf{Conclusion.}
In this paper we have characterised the equilibrium properties of the $\theta$--angled spin--boson model, in the quantum and classical regime.
\blue{Firstly, for the classical case, we have derived a compact analytical expression for the equilibrium state of the spin, that is valid at arbitrary coupling to the harmonic reservoir.
This is of great practical relevance as it allows to  analytically obtain all equilibrium properties of the spin at any coupling strength.
It remains an open question \cite{trushechkin2021b} to find a similarly general analytical expressions for the quantum case.}
\blue{Secondly, we have proved that the quantum \MFGS{} partition function, including environmental terms, converges to its classical counterpart in the large-spin limit at all coupling strengths. Our results provide direct insight in the difference between quantum and classical states of a spin coupled to a noisy environment. Apart from being of purely fundamental interest, this will constitute key information for many quantum technologies \cite{Auffeves2022}, and ultimately links to the quantum supremacy debate.
}

\blue{Third, a large and growing body of literature identifies coherences as quantum signatures and attributes speed-ups, e.g. in quantum computing and quantum biology, or efficiency gains, e.g. in quantum thermodynamics, to quantum coherences. 
Here we demonstrated that even the equilibrium states of classical open spins host `coherences' when the environment couples asymmetrically. Thus, measures other than `coherences' may be required to certify the quantum origin of certain speed-ups or efficiency improvements in the future.
Finally, we presented the first quantitative characterisation of the coupling parameter values that put the spin-boson model in the ultraweak, weak, intermediate, or ultrastrong coupling regime, both for the quantum case as well as the classical setting. This classification will be important in many future studies of the spin-boson model, quantum or classical, for which it provides the tool to chose the correct approximation for a specific parameter set. 
}

\medskip

\blue{
\noindent\textbf{Code availability.}
The code used to produce Fig.~\ref{fig:regimes} and Fig.~\ref{fig:clregimes} is publicly available online at \href{https://github.com/quantum-exeter/SpinMFGS}{https://github.com/quantum-exeter/SpinMFGS}. It can be used to make analogous plots for a desired coupling angle $\theta$ and  spin length $\SO$.}

\medskip


\noindent\textbf{Acknowledgments.}
We thank Anton Trushechkin for stimulating discussions on the subject of this research.
FC gratefully acknowledges funding from the Foundational Questions Institute
Fund (FQXi-IAF19-01).
SS is supported by a DTP grant from EPSRC (EP/R513210/1).
SARH acknowledges funding from the Royal Society and TATA (RPG-2016-186).
JA, MB and JC gratefully acknowledge funding from EPSRC (EP/R045577/1).
JA thanks the Royal Society for support.



%



\newcommand{\refeqtaumf}{\eqref{eq:taumf}}
\newcommand{\refeqHtot}{\eqref{eq:Htot}}
\newcommand{\refeqHint}{\eqref{eq:Hint}}
\newcommand{\refeqreorg}{\eqref{eq:reorg}}
\newcommand{\refeqrelstrdef}{\eqref{eq:relstrdef}}
\newcommand{\reffigcss}{\ref{fig:css}}
\newcommand{\reffigeqmagvsT}{\ref{fig:eqmag_vs_T}}
\newcommand{\reffigregimes}{\ref{fig:regimes}}


\appendix

\renewcommand\thefigure{S\arabic{figure}}    
\setcounter{figure}{0}


\section{Tracing for spin and reservoir, in classical and quantum setting}
\label{sec:tracing}


\subsection{Spin tracing in the classical setting}

For a classical spin of length $\SO$, with components $\Sx,\Sy, \Sz$, one can
change into spherical coordinates, i.e.
\begin{align} \label{eq:sm:spinreplacements}
  \Sx &= \SO\sin\vartheta\cos\varphi, \quad
  \Sy = \SO\sin\vartheta\sin\varphi, \\ \nonumber 
  \Sz &= \SO\cos\vartheta,
  \qquad \vartheta\in[0,\pi],\varphi\in[0,2\pi].
\end{align}
Then, traces of functions $A(\Sx,\Sz)$ are evaluated as
\begin{align} \label{eq:sm:classspintrace}
    \trS^\class[A(\Sx,\Sz)] &= \\ \nonumber
    \frac{1}{4\pi} \IntPhi\IntTheta &\sin\vartheta
        A(\SO\sin\vartheta\cos\varphi, \SO\cos\vartheta).
\end{align}


\subsection{Spin tracing in the quantum setting}

For a quantum spin $\SO$, given any orthogonal basis $\ket{m}$, then the trace
of functions of the spin operators $A(\Sx,\Sz)$ are evaluated as
\begin{equation}
  \trS^\quant[A(\Sx,\Sz)] = \sum_{m}\bra{m}A(\Sx,\Sz)\ket{m}.
\end{equation}


\subsection{Reservoir traces}

When taking traces over the environmental degrees of freedom (in either the
classical or quantum case), we ought to first discretise the energy spectrum of
$\HR$. This is because, strictly speaking, the partition function for the
reservoir, $\ZR = \tr[\exp(-\beta\HR)]$, is not well defined in the continuum
limit. Thus, we write
\begin{equation}
  \HR = \sum_{n=0}^{\infty}
    \frac{1}{2}\left(\Pwn^2 + \omega_n^2 \Xwn^2\right).
\end{equation}
Then, for example, the classical partition function of the environment is
\blue{
\begin{equation}
  \ZRcl = \prod_{n=0}^{\infty}\left[
  \int_{-\infty}^{+\infty}\hspace{-1em}\d\Xwn
  \int_{-\infty}^{^\infty}\hspace{-1em}\d\Pwn 
  e^{-\frac{\beta}{2}\left(\Pwn^2 + \omega_n^2 \Xwn^2\right)}\right],
\end{equation}
}
and similarly for the quantum case.


\section{Expectation values from the partition function}
\label{sec:expvaluespf}

With the partition function of the \MFGS{} we can proceed to calculate the $\Sz$
and $\Sx$ expectation values as follows.


\subsection{Classical case}

For the classical spin, from~\refeqtaumf{} we have the partition function
\begin{align}
    \ZSMFcl &= \\ \nonumber
    \frac{1}{4\pi} &\IntPhi\IntTheta \sin\vartheta
    e^{-\beta(-\wL\Sz(\vartheta,\varphi) - \Q\Stheta^2(\vartheta,\varphi))}.
\end{align}
While obtaining the $\Sz$ expectation value is straightforward, the $\Sx$ case
may seem less obvious. It is therefore convenient to do a change of coordinates
\begin{align}
    \Sabs_{z'}(\vartheta,\varphi) &= \Sz(\vartheta,\varphi)\cos\theta - \Sx(\vartheta,\varphi)\sin\theta,
    \\ \nonumber 
    \Sabs_{x'}(\vartheta,\varphi) &= \Sx(\vartheta,\varphi)\cos\theta + \Sz(\vartheta,\varphi)\sin\theta.
\end{align}
Defining $h_{x'} = -\wL\sin\theta$,
$h_{z'} = -\wL\cos\theta$, we then have that
\begin{align}
    \ZSMFcl = \frac{1}{4\pi}&\IntPhi\IntTheta\sin\vartheta
    \\ \nonumber
        &e^{-\beta(h_{z'}\Sabs_{z'}(\vartheta,\phi)
            + h_{x'}\Sabs_{x'}(\vartheta,\phi)
            - \Q\Sabs_{z'}^2(\vartheta,\phi))},
\end{align}
and we can obtain the $\Sabs_{z'}$ and $\Sabs_{x'}$ expectation values as usual,
i.e.
\begin{equation} \label{eq:sm:expvalfromZ}
    \expval{\Sabs_{x',z'}} = -\frac{1}{\beta}\frac{\partial}{\partial h_{x',z'}}
        \log\ZSMFcl.
\end{equation}
Finally, by linearity, we have that
\begin{align} \label{eq:sm:expvalfromZ2}
    \expval{\Sx} &= \expval{\Sabs_{x'}}\cos\theta - \expval{\Sabs_{z'}}\sin\theta,
    \\ \nonumber 
    \expval{\Sz} &= \expval{\Sabs_{z'}}\cos\theta + \expval{\Sabs_{x'}}\sin\theta.
\end{align}


\subsection{Quantum case}

For the quantum case we proceed in a completely analogous manner. We have that
\begin{align}
    \expval{\Sx} &= \trqu\left[\Sx e^{-\beta(-\wL\Sz + \Stheta\Bop + \HR)}\right],
    \\ \nonumber 
    \expval{\Sz} &= \trqu\left[\Sz e^{-\beta(-\wL\Sz + \Stheta\Bop + \HR)}\right].
\end{align}
Starting from the partition function
\begin{equation}
    \ZSRqu = \trqu\left[e^{-\beta(-\wL\Sz + \Stheta\Bop + \HR)}\right],
\end{equation}
we define a new set of rotated operators,
\begin{align}
    \Sabs_{z'} &= \Sz\cos\theta - \Sx\sin\theta,
    \\ \nonumber 
    \Sabs_{x'} &= \Sx\cos\theta + \Sz\sin\theta,    
\end{align}
and variables $h_{x'} = -\wL\sin\theta$, $h_{z'} = -\wL\cos\theta$, so that
\begin{equation}
    \ZSRqu
    = \trqu[e^{-\beta(h_{z'}\Sabs_{z'} + h_{x'}\Sabs_{x'}
        + \Sabs_{z'}\Bop + \HR)}].
\end{equation}
Then, we proceed in an analagous way as in \eqref{eq:sm:expvalfromZ}
and \eqref{eq:sm:expvalfromZ2}.


\subsection{Example: Ultrastrong limit}

Let us consider the quantum ultrastrong partition function,
\begin{equation}
    \ZQMFUS = \cosh\left(\beta\wL\SO\cos\theta\right).
\end{equation}
Following the procedure outlined above, we have that $h_{z'} = -\wL\cos\theta$,
and therefore
\begin{equation}
    \ZQMFUS = \cosh\left(\beta\SO h_{z'}\right).
\end{equation}
Therefore, the expectation values of the transformed observables are
\begin{align}
    \expval{\Sabs_{x'}}
        &= -\frac{1}{\beta}\frac{\partial}{\partial h_{x'}} \log\ZQMFUS
        = 0,
    \\
    \expval{\Sabs_{z'}}
        &= -\frac{1}{\beta}\frac{\partial}{\partial h_{z'}} \log\ZQMFUS
        = -\SO\tanh\left(\beta\SO h_{z'}\right)
        \nonumber \\
        &= -\SO\tanh\left(\beta\wL\SO\cos\theta\right).
\end{align}
Therefore, in the original variables we have
\begin{align}
    \expval{\Sx}
    &= \SO\sin\theta\tanh\left(\beta\wL\SO\cos\theta\right),
    \\
    \expval{\Sz}
    &= -\SO\cos\theta\tanh\left(\beta\wL\SO\cos\theta\right),
\end{align}
in agreement with what is later obtained in \supp~\suppus~directly
from the \MFGS{} in the ultra-strong limit.


\section{Derivation of classical \MFGS{} state for arbitrary coupling}
\label{sec:classicaltmf}

In this section we derive the mean force Gibbs state of the classical spin
for arbitrary coupling strength. As discussed in~\ref{sec:tracing}, we
discretise the environmental degrees of freedom, and thus we have for the total
Hamiltonian,~\refeqHtot
\begin{align}
  \Htot = -\wL\Sz + \sum_{n=0}^{\infty}
    \Bigg[\frac{1}{2}\big(\Pwn^2 &+ \omega_n^2 \Xwn^2\big)
    \nonumber \\ &+ \Stheta\Cwn\Xwn \Bigg].
\end{align}
On `completing the square', we get
\begin{align}
  \Htot =
  -\wL\Sz + \sum_{n=0}^{\infty}
    \frac{1}{2}\Bigg[\Pwn^2 &+ \omega_n^2
    \Big(\Xwn - \frac{\Stheta\Cwn}{\omega_n^2}\Big)^2
    \nonumber \\ &- \frac{(\Stheta\Cwn)^2}{2\omega_n^2}\Bigg].
\end{align}
The partition function is then,
\begin{equation}
  \ZSRcl =\blue{\frac{1}{4\pi}} \IntPhi\IntTheta\sin\vartheta e^{-\beta\Heff} \ZRcl.
\end{equation}
Here, there appears an effective system Hamiltonian given by
\begin{equation}
  \Heff \equiv -\wL\Sz - \Q\Stheta^2
\end{equation}
where the reorganization energy $\Q$, is given by Eq.~\refeqreorg{} of the
main text, and
\begin{align}
  \ZRcl = \prod_n
  \int_{-\infty}^\infty\vspace{-1em}\d\Xwn &
  \int_{-\infty}^{^\infty}\vspace{-1em}\d\Pwn 
  \\ \nonumber
  &e^{-\frac{1}{2}\beta\big(\Pwn^2 + \omega_n^2
    \big(\Xwn - \frac{\Stheta\Cwn}{\omega_n^2}\big)^2\big)},
\end{align}
is the partition function for the reservoir only.
\blue{
Note that, despite seemingly depending on the spin coordinates, this
last integral coincides with the reservoir partition function since once
one carries out the Gaussian integral, the dependence on $\Stheta$ vanishes.
}

While it is possible to derive an expression for $\ZR$, its details are not
needed as it depends solely on reservoir variables and can be divided out to
yield the system's \MFGS{} partition function,
\begin{align} \label{eq:sm:ZS}
  \ZSMFcl &= \frac{\ZSRcl}{\ZRcl} =\frac{1}{4\pi} \IntPhi\IntTheta\sin\vartheta e^{-\beta\Heff}, 
\end{align}
where $\Heff$ includes all spin terms independent of the coordinates of the
environment. Finally, the \MFGS{} is given by
\begin{equation} 
  \tsysMFGS = \frac{1}{\ZSMFcl} e^{-\beta\Heff},
\end{equation}
which is precisely Eq.~\refeqtaumf{} of the main text.


In terms of polar coordinates, we have $\Sz = \SO\cos\vartheta$ and $\Sx =
\SO\cos\varphi\sin\vartheta$. Therefore, $\Stheta =
\SO\left(\cos\vartheta\cos\theta - \sin\vartheta\cos\varphi\sin\theta\right)$,
and we have that
\begin{align}
  \Heff(\vartheta,\varphi) = &-\wL\SO\cos\vartheta
    \\ \nonumber
    &- \SO^2\Q\left(\cos\theta\cos\vartheta
     - \sin\theta\sin\vartheta\cos\varphi\right)^2.
\end{align}
The equilibrium state of the spin is then entirely determined by $\ZSMFcl$. The
classical expectation values for the spin components $\Sz$ and $\Sx$ are then
given by
\begin{align}
  \expval{\sz} &= \frac{\expval{\Sz}}{\SO}
  \label{eq:sm:Sz}
  \\ \nonumber
  &= \frac{1}{\ZSMFcl} \IntPhi\IntTheta \sin\vartheta\cos\vartheta
    e^{-\beta\Heff(\vartheta,\varphi)}
  \\
  \expval{\sx} &= \frac{\expval{\Sx}}{\SO}
  \label{eq:sm:Sx}
  \\ \nonumber
  &= \frac{1}{\ZSMFcl} \IntPhi\cos\varphi \IntTheta\sin^2\vartheta
    e^{-\beta\Heff(\vartheta,\varphi)}.
\end{align}
The integral expressions for the expectation values above cannot in general be
expressed in a closed form, but can be readily evaluated numerically for
arbitrary coupling strength $\Q$.


\section{Quantum-classical correspondence for the \MF{} partition functions}
\label{sec:quantumtoclassical} 

Starting from equation \eqref{eq:sm:13} of the main text, we
now ``complete the square'' for the combination
\begin{equation}
  \hR + \stheta\IntW\Cw\sqrt{\SO}\xw, 
\end{equation}
to arrive at
\begin{align}
  &\frac{1}{2}\IntW\left(\pw^2 + \omega^2\left(\xw + \stheta
    \frac{\Cw\sqrt{\SO}}{\omega^2}\right)^2\right)
  - \stheta^2\SO\Q(\SO)
  \nonumber \\
  &=
  \hRshift - \stheta^2\SO\Q(\SO),
\end{align}
where $\Q(\SO) = \IntW\Cw^2(\SO)/(2\omega^2)$ is the reorganisation
energy, see~\refeqreorg{}. Note that because of the scaling $\Cw \propto
1/\sqrt{\SO}$, the product $\SO\Q(\SO) = \alp \, \wL$ is independent of $\SO$.
Here, we have defined the reservoir Hamiltonian 
\begin{equation}
  \hRshift = \frac{1}{2}\IntW\left(\pw^2 + \omega^2\left(\xw +
    \stheta\frac{\Cw\sqrt{\SO}}{\omega^2}\right)^2\right),    
\end{equation}
where the oscillator centres have been shifted.

Applying~\eqref{eq:sm:31} to the total spin-reservoir Hamiltonian $\Htot$, 
and immediately taking the reservoir trace on both sides, gives 
\begin{align}
  &\lim_{\SO\to\infty} \frac{\hbar}{2\SO+\hbar} \ZSRqu(\beta,\SO,\alp) \nonumber \\
  = &\lim_{\SO\to\infty} \frac{\hbar}{2\SO+\hbar}
    \trSRqu\left[e^{-\beta'\left(-\wL\sz + \hRshift -
    \stheta^2 \, \alp \, \wL\right)}\right] \nonumber \\
  = &\lim_{\SO\to\infty}\frac{1}{4\pi}\IntPhi\IntTheta\sin\vartheta
    \nonumber \\
    &\qquad\trRqu\left[e^{-\beta'\left(-\wL\cos\vartheta + \hRshift -
    \stheta^2(\vartheta,\phi')\, \alp\, \wL\right)}\right] \nonumber \\ 
  = &\lim_{\SO\to\infty}\frac{1}{4\pi}\IntPhi\IntTheta\sin\vartheta
    \nonumber \\
    &\qquad e^{-\beta'\left(-\wL\cos\vartheta - \stheta^2(\vartheta,\phi')\, \alp\, \wL\right)}
    \trRqu\left[e^{-\beta'\hRshift}\right],
  \label{eq:sm:new33}
\end{align}
where the trace over the reservoir now factors out and
\begin{equation}
  \stheta(\vartheta,\phi') = \cos\theta\cos\vartheta -
    \sin\theta\cos\varphi\sin\vartheta.
\end{equation}
The reservoir trace factor gives 
\begin{align}
  \trRqu\left[e^{-\beta'\hRshift}\right] 
  &= \trRqu\left[e^{-\beta\frac{1}{2}
    \IntW \left(\Pw^2 + \omega^2\left(\Xw + \mu_\omega\right)^2\right)}\right]
  \nonumber \\
  &= \ZRqu(\beta),
\end{align}
with $\mu_\omega = \Stheta\frac{\Cw}{\omega^2}$ a shift in the centre 
position of the oscillators.
The operators $\Xw + \mu_\omega$ have the same commutation relations
with the $\Pw$ as the $\Xw$ themselves. Thus such a shift does not
affect the trace and the result is the bare quantum reservoir partition 
function at inverse temperature $\beta$, i.e. $\ZRqu(\beta)$. 

Dividing by $\ZRqu(\beta)$ on both sides, putting it all together, we find
\begin{align}
  &\lim_{\SO\to\infty} \frac{\hbar}{2\SO+\hbar} \frac{\ZSRqu(\beta,\SO, \alp)}{\ZRqu(\beta)}
  \label{eq:sm:new34} \\
  &= \frac{1}{4\pi}\IntPhi\IntTheta\sin\vartheta \, e^{-\beta'\left(-\wL\cos\vartheta - \stheta^2(\vartheta,\phi')\, \alp\, \wL\right)}, \nonumber
\end{align}
where we have dropped the limit symbol since there is no dependence on $\SO$.

Now we may replace again $\beta' = \beta\SO$, and the RHS emerges
as the spin's classical mean force partition function
$\ZSMFcl(\beta\SO, \alp)$ cf.~\refeqtaumf{}, where the classical trace
is taken according to~\eqref{eq:sm:classspintrace}.
Moreover, the fraction of total quantum partition function divided by bare
reservoir partition function is the quantum mean force partition
function~\cite{campisi2009,strasberg2020}.
Thus, we conclude: 
\begin{align} \label{eq:sm:new35:smver}
  &\lim_{\SO\to\infty} \frac{\hbar}{2\SO+\hbar} \, \ZSMFqu(\beta,\SO, \alp) 
  \nonumber \\
  &= \lim_{\SO\to\infty} \frac{\hbar}{2\SO+\hbar} \frac{\ZSRqu(\beta,\SO, \alp)}{\ZRqu(\beta)}
  = \ZSMFcl(\beta\SO, \alp).
\end{align}


\section{Quantum Reaction Coordinate mapping}
\label{sec:qrc}

The Reaction Coordinate mapping method
\cite{iles-smith2014,iles-smith2016,nazir2018,correa2019}
is a technique for dealing with systems strongly coupled to bosonic
environments. To do so, it isolates a single collective environmental
degree of freedom, the so called ``reaction coordinate'' (RC), that interacts
with the system via an effective Hamiltonian. The rest of the environmental
degrees of freedom manifest as a new bosonic environment coupled to the RC.
Concretely, for our total Hamiltonian \refeqHtot{}, the effective Hamiltonian
that we have to consider is
\begin{equation}
  \HRCtot = \HS + \HRC + \HRCint + \HResint + \HRes,
\end{equation}
where $\HRC$ is the Hamiltonian of the RC mode,
\begin{equation}
  \HRC = \hbar\rcfreq a^\dagger a,
\end{equation}
with $a^\dagger$ the creation operator of a quantum harmonic oscillator of
frequency $\rcfreq$; $\HRCint$ is the spin--RC interaction
\begin{equation}
  \HRCint = \rcint\Stheta(a + a^\dagger),
\end{equation}
where $\rcint$ determines the the coupling strength between the RC mode and
the spin; $\HRes = \int\d\omega(\pw^2 + \omega^2\qw^2)/2$ is the Hamiltonian of
the residual bosonic bath; and finally the residual bath-RC interaction
$\HResint$ is
\begin{equation}
  \HResint = (a + a^\dagger)\IntW\sqrt{2\omega\Jrc}\qw,
\end{equation}
with $\Jrc$ the spectral density of the residual bath.

Given $\Htot$, for an appropriate choice of $\Jrc$ (which depends on the
original Hamiltonian spectral density and coupling), it has been proven that
the reduced dynamics of the spin under $\Htot$ are exactly the same as those of
the spin under the effective Hamiltonian $\HRCtot$ \cite{nazir2018}.
In general, the mapping between the original spectral density, $\Jw$, and that
of the RC Hamiltonian, $\Jrc$, is hard to find.
However, one particular case were there is a simple closed form for $\Jrc$ is
that of a Lorentzian spectral density $\Jw$ (see main text).
%
%
In such case, the $\Jrc$ spectral density is exactly Ohmic 
\cite{iles-smith2014,iles-smith2016,nazir2018}, i.e. has the form
\begin{equation} \label{eq:sm:Johm}
  \Jrc = \rcdis\omega e^{-\omega/\omega_{\mathrm{c}}}, \qquad
    \omega_{\mathrm{c}} \to \infty.
\end{equation}
Furthermore, the RC effective Hamiltonian parameters ($\rcfreq$, $\rcint$ and
$\rcdis$) are given in terms of the Lorentzian parameters by
\begin{align}
  \rcfreq &= \omega_0, \\
  \rcint &= \sqrt{\Q\omega_0}, \\
  \rcdis &= \frac{\Gamma}{2\pi\omega_0}.
\end{align}
Noticeably, by appropriately choosing $\Q$, $\Gamma$, and $\omega_0$, we
can have an initial Hamiltonian with arbitrarily strong coupling to the full
environment (i.e. arbitrarily strong $\Q$), while having arbitrarily small
coupling to the residual bath of the RC Hamiltonian (i.e. arbitrarily small
$\rcdis$).

As mentioned, it has been shown that the reduced spin dynamics under $\Htot$ with
Lorentzian spectral density (see main text) is exactly the same as the reduced spin
dynamics under $\HRCtot$ with spectral density \eqref{eq:sm:Johm}.
In particular, the steady state of the spin will also be the same.
Therefore, it is reasonable to expect that the spin \MF{} state obtained with
$\Htot$ will be the same as the spin \MF{} state for $\HRCtot$, i.e.
\begin{equation}
  \tsysMFGS
  = \ZSMF^{-1} \trR[e^{-\beta\Htot}]
  = \ZSMF'^{-1} \trR[e^{-\beta\HRCtot}].
\end{equation}

We now assume that $\rcdis$ is arbitrarily small, so that the \MF{} state is
simply going to be given by the Gibbs state of spin+RC, i.e.
\begin{align} \label{eq:sm:taumfapprox}
  \tsysMFGS &= \ZSMF'^{-1} \trR[e^{-\beta\HRCtot}]
  \\ \nonumber
  &\approx \ZSMF''^{-1}
  \trR\Big[e^{-\beta\left(\HS + \rcfreq a^\dagger
    a + \rcint\Stheta(a+a^\dagger)\right)}\Big].
\end{align}
It is key here to observe that the condition $\rcdis \to 0$ does not imply any
constraint on the coupling strength to the original environment, since we can
always choose $\Gamma$ and $\omega_0$ so that $\rcdis$ is arbitrarily small,
while allowing $\Q$ to be arbitrarily large.

Finally, to numerically obtain the \MF{} state, since unfortunately
\eqref{eq:sm:taumfapprox} does not have a general closed form, we numerically
evaluate \eqref{eq:sm:taumfapprox} by diagonalising the full Hamiltonian and
then taking the partial trace over the RC.
To numerically diagonalise this Hamiltonian we have to choose a cutoff on the
number of energy levels of the RC harmonic oscillator.
This cutoff was chosen by increasing the number of levels until observing
convergence of the numerical results.


\section{Quantum to classical limit in the weak coupling approximation}
\label{sec:quantumweakcoupling}

In this section we
explicitly compute
the large spin limit for the weak coupling expressions of the
classical and quantum mean force Gibbs states.
These results are used in the characterisation of the different coupling regimes.

\bigskip

Since we are going to perform perturbative expansions in the coupling strength,
in what follows we introduce, for book-keeping purposes, an adimensional
parameter $\lambda$ in the interaction, so that $\Hint$ now reads
\begin{equation}
  \Hint = \lambda\Stheta\IntW\Cw\Xw.
\end{equation}
This will allow us to properly keep track of the order of each therm in the
expansion. Finally, at the end of the calculations we will take $\lambda = 1$.


\subsection{Classical spin: weak coupling}

Here, we derive the classical weak coupling expectation values starting from the
exact \MFGS{} found in \ref{sec:classicaltmf}. The effective Hamiltonian, with
the inclusion of the parameter $\lambda$ now reads
\begin{equation}
  \Heff = -\wL\Sz - \lambda^2\Q\Stheta^2.
\end{equation}
For weak coupling, the expressions for $\ZSMFcl,\expval{\Sz}$ and $\expval{\Sx}$
can be approximated by treating the term $\lambda^2\SO^2\Q$ as a perturbation.
Therefore, expanding $\exp[-\beta\Heff]$ to lowest order in $\lambda$ we have
\begin{align}
  &e^{-\beta\Heff} = e^{\beta\wL\SO\cos\vartheta}
      \Big[1 +
  \\ \nonumber
      &\beta\lambda^2\SO^2\Q \left(\cos\theta\cos\vartheta -
      \sin\theta\sin\vartheta\cos\varphi\right)^2\Big]
      + \order(\lambda^4),
\end{align}
from which we can determine the weak coupling limit of the classical spin
partition function and spin expectation values.


\subsubsection{Standard Gibbs results for a classical spin}

First, here we write the partition function and spin expectation values for a
classical spin in the standard Gibbs state for the bare Hamiltonian $\HS$ (i.e.
in the limit of vanishing coupling, $\lambda = 0$). These expressions will be
useful to later on to write the second order corrections.

For the partition function we have that
\begin{align}
  \ZO^\text{cl}
  &= \frac{1}{4\pi}\IntPhi\IntTheta\sin\vartheta 
    \exp[\beta\wL\SO\cos\vartheta]
  \nonumber \\
  &= \frac{\sinh(\beta\wL\SO)}{\beta\wL\SO}.
\end{align}

For $\Sx$ we have that is trivially $0$ by symmetry, i.e.
\begin{align}
  \expval{\Sx}_0
  &= \frac{1}{\ZO^\text{cl}} \SO \IntPhi\cos\varphi
    \IntTheta \sin^2\vartheta e^{\beta\wL\SO\cos\vartheta}
  \\ \nonumber
  &= 0.
\end{align}

For the expectation value of the powers of $\Sz$
(which will be useful later), we have
\begin{equation}
  \expval{\Sz^n}_0 = \frac{2\pi}{\ZO^\text{cl}}\SO^n
    \IntTheta\sin\vartheta\cos^n\vartheta \,
    e^{\beta\wL\SO\cos\vartheta}.
\end{equation}
In particular, we find
\begin{align}
    \label{eq:sm:ZerothOrder}
    \expval{\Sz}_0 &= \SO\coth(\beta\wL\SO) - \frac{1}{\beta\wL}, \\
    \expval{\Sz^2}_0 &= \SO^2 - \frac{2\SO\coth(\beta\wL\SO)}{\beta\wL} + \frac{2}{(\beta\wL)^2}, \\
    \expval{\Sz^3}_0 &= \SO^3\coth(\beta\wL\SO) - \frac{3\SO^2}{\beta\wL}
    \nonumber \\
    &+\frac{6\SO\coth(\beta\wL\SO)}{(\beta\wL)^2} - \frac{6}{(\beta\wL)^3}.
\end{align}


\subsubsection{Classical spin partition function for weak coupling}

Expanding the partition function to second order in $\lambda$ we find that
\begin{align}
  \ZSMFcl =
  &\frac{1}{4\pi}\IntPhi\IntTheta\sin\vartheta
  \Bigg[e^{\beta\wL\SO\cos\vartheta} 
  \nonumber \\
  &+ \frac{\beta\lambda^2\SO^2\Q}{4\pi}
    e^{\beta\wL\SO\cos\vartheta} \big(\cos\theta\cos\vartheta
  \nonumber \\
  &\qquad - \sin\theta\sin\vartheta\cos\varphi\big)^2\Bigg]
  + \order(\lambda^4).
\end{align}
The first term can be recognised as the partition function for the
bare system, $\ZO^\text{cl}$. The $\varphi'$ integral in the second
term is straightforward to perform,
\begin{align}
  \ZSMFcl &= \ZO^\text{cl}
  \nonumber \\
  &+ \frac{1}{2}\beta\lambda^2\SO^2\Q\IntTheta\sin\vartheta
  e^{\beta\wL\SO\cos\vartheta}
  \nonumber \\
  &\qquad\qquad\left[(3\cos^2\theta-1)\cos^2\vartheta + \sin^2\theta\right]
  \nonumber \\
  &+ \order(\lambda^4).
\end{align}
We typically require the inverse of the partition function, which to lowest
order in the perturbation is
\begin{align}
  \ZSMFcl &= (\ZO^\text{cl})^{-1}\Bigg[
  \nonumber \\
  &1 - \pi\beta\lambda^2\SO^2\Q\ZO^{-1}\IntTheta\sin\vartheta
  e^{\beta\wL\SO\cos\vartheta}
  \nonumber \\
  &\qquad\qquad
  \left((3\cos^2\theta-1)\cos^2\vartheta + \sin^2\theta\right)\Bigg]
  \nonumber \\
  &+ \order(\lambda^4).
\end{align}



Now turning to the expectation value $\expval{\Sz}$, given in
Eq.~\eqref{eq:sm:Sz}, and carrying out the same lowest order expansion
we get
\begin{align}\label{eq:sm:SzClassical1}
  \expval{\Sz} &= \expval{\Sz}_0
  \nonumber \\
  &+ \frac{1}{2}\beta\lambda^2\Q\left[(3\cos^2\theta-1) (\expval{\Sz^3}_0 -
    \expval{\Sz}_0 \expval{\Sz^2}_0)\right]
  \nonumber \\
  &+ \order(\lambda^4).
\end{align}
This result will be compared later to the quantum weak coupling result 
obtained in the large spin (classical) limit.


\subsubsection{Classical \texorpdfstring{$\expval{\Sx}$}{<Sx>} for weak coupling}

A similar calculation can be followed for $\expval{\Sx}$, the main difference
being in the handling of the $\varphi'$ integral. Thus, we find
\begin{equation} \label{eq:sm:SxClassical1}
  \expval{\Sx} = -\tfrac{1}{2}\sin2\theta\beta\lambda^2\Q
      \left(\expval{\Sz}_0\SO^2 - \expval{\Sz^3}_0\right) + \order(\lambda^4),
\end{equation}
where we have used that $\ZO/\ZSMFcl = 1$ to lowest order.

Using the zeroth order expressions for $\expval{\Sz}_0$ and $\expval{\Sz^3}_0$
from \eqref{eq:sm:ZerothOrder} we get the result in terms of the scaled
temperature $\beta' = \beta\SO$
\begin{align}\label{eq:sm:SxClassical2}
  \expval{\sx} = -\frac{\sin2\theta\lambda^2\SO\Q}{\wL}
      \Bigg(&1-\frac{3\coth(\beta'\wL)}{\beta'\wL}
      \\ \nonumber
      &+ \frac{3}{(\beta'\wL)^2}\Bigg) + \order(\lambda^4).
\end{align}
This result will be compared later to the quantum weak coupling result obtained
in the large spin (classical) limit.


\subsection{Quantum spin: weak coupling}

In general, the quantum mean force Gibbs state is given by
\begin{equation} \label{eq:sm:QMFGS}
  \tsysMFGS = \frac{\trRqu\left[e^{-\beta\Htot}\right]}{\Ztotqu},
\end{equation}
with $\Htot$ given by Eq.~\refeqHtot{} of the main text. Unfortunately,
determining the form of $\tsysMFGS$ and the various expectation values for the
spin components is unfeasible in the general case, but limiting forms are
available.
Here we derive the spin expectation values in the weak coupling limit, and then
later on take the large spin limit to explicitly verify the quantum-to-classical
transition.


\subsubsection{Standard Gibbs results for a quantum spin}

Here, we present the results of the standard Gibbs state for the quantum spin
(i.e. in the limit of vanishing coupling, $\lambda = 0$). The Gibbs state for
the system's bare Hamiltonian is
\begin{equation}
  \tauS = \frac{e^{\beta\wL\Sz}}{\ZSqu}, \quad
  \ZSqu = \tr\left[e^{\beta\wL\Sz}\right].
\end{equation}
We also have that $[\tauS,\Sz] = 0$. The trace is readily evaluated, yielding
the partition function
\begin{equation}
  \ZO^\text{qu} = \frac{\sinh\beta\wL(\SO + \tfrac{\hbar}{2})}{\sinh\tfrac{\hbar}{2}\beta\wL},
\end{equation}
from which we can derive the expectation values of $\Sz$, $\Sz^2$ and $\Sz^3$,
\begin{equation}
  \expval{\Sz^n}_0 = \frac{1}{\ZO^\text{qu}} \frac{\d^n}{\d(\beta\wL)^n}\ZO^\text{qu}.
\end{equation}
We find,
\begin{align}
\label{eq:sm:ZerothOrderQuantumExpectationValues}
  \expval{\Sz}_{0} &=
    (\SO + \tfrac{\hbar}{2})\coth(\beta\wL(\SO + \tfrac{\hbar}{2}))
    - \tfrac{\hbar}{2}\coth(\tfrac{\hbar}{2}\beta\wL) \\
  \expval{\Sz^2}_0 &=
    (\SO + \tfrac{\hbar}{2})^2
    \nonumber \\
    &- \hbar(\SO + \tfrac{\hbar}{2})\coth(\tfrac{\hbar}{2}\beta\wL)\coth(\beta\wL(\SO
      + \tfrac{\hbar}{2}))
    \nonumber \\
    &+ \tfrac{\hbar^2}{4}\left(2\coth^2(\tfrac{\hbar}{2}\beta\wL) - 1\right) \\
  \expval{\Sz^3}_0 &=
    (\SO + \tfrac{\hbar}{2})^3\coth(\beta\wL(\SO + \tfrac{\hbar}{2}))
    \nonumber \\
    &- \tfrac{3\hbar}{2}(\SO + \tfrac{\hbar}{2})^2 \coth(\tfrac{\hbar}{2}\beta\wL)
    \nonumber \\
    &+ \tfrac{3\hbar^2}{4}(\SO + \tfrac{\hbar}{2})\coth(\beta\wL(\SO + \tfrac{\hbar}{2})) 
    \nonumber \\
    &\qquad\qquad\left(2\coth^2(\tfrac{\hbar}{2}\beta\wL) - 1\right)
    \nonumber \\
    &- \tfrac{3\hbar^3}{4}\coth^3(\tfrac{\hbar}{2}\beta\wL)
    + \tfrac{5\hbar^3}{8}\coth(\tfrac{\hbar}{2}\beta\wL).
\end{align}

\subsubsection{General form of weak coupling density operator}

For a total Hamiltonian $\HS + \HR + \Hint$ with
interaction of the form $\Hint = \lambda\Xop\Bop$,
the general expression for the \emph{unnormalised} mean
force state to second order in the interaction is given by \cite{cresser2021a}
\begin{align}\label{eq:sm:8}
  &\rhowkun = \tauS
  \nonumber \\
  &+ \lambda^2\sum_{n}\left(\left[\Xop_n^\dagger,\tauS\Xop_n\right]A'_\beta(\omega_n)
    + \beta\tauS\Xop_n\Xop_n^\dagger\,A_\beta(\omega_n)\right)
  \nonumber \\
  &+ \lambda^2\sum_{m \ne n}\omega_{mn}^{-1}\left(\left[\Xop_m,\Xop_n^\dagger\tauS\right]
    + \left[\tauS\Xop_n,\Xop_m^\dagger\right]\right)\,A_\beta(\omega_n),
\end{align}
where the system operator $\Xop$ is expanded in terms of the energy
eigenoperators $\Xop_n$
\begin{equation}
  \Xop = \sum_{n} \Xop_n,
\end{equation}
with $\left[\HS,\Xop_n\right] = \omega_n\Xop_n$, and $\omega_n$ are Bohr
frequencies for the system. We have $\Xop_n^\dagger = \Xop_{-n}$ and $\omega_n =
-\omega_{-n}$. The quantities $A_\beta(\omega_n)$ are determined by the
correlation properties of the reservoir operator $\Bop$ and are given by
\begin{align} \label{eq:sm:9}
  A_\beta(\omega_n)
  &= \IntW \Jw\left(\frac{n_\beta(\omega)+1}{\omega-\omega_n}
    - \frac{n_\beta(\omega)}{\omega+\omega_n}\right), \\
  A'_\beta(\omega_n)
  &= \IntW \frac{\Jw}{\hbar}\left(\frac{n_\beta(\omega)+1}{(\omega-\omega_n)^2}
    + \frac{n_\beta(\omega)}{(\omega+\omega_n)^2}\right).
\end{align}
We can separate out the particular case of $\omega_n = 0$, for which we find
\begin{equation} \label{eq:sm:13}
  A_\beta(0) = \IntW\frac{\Jw}{\omega} = \Q.
\end{equation}

It turns out that we will require various symmetric and antisymmetric
combinations of $A_\beta(\omega_n)$ and $A'_\beta(\omega_n)$.
[Note that in the following (and in the initial definition of the quantity
$A'_\beta(\omega_n)$), the dash indicates a derivative wrt to the argument
$\omega_n$. Thus the quantity $A_\beta'(-\omega_n)$ is a derivative wrt
$-\omega_n$, i.e., $A_\beta'(-\omega_n) = -\d A_\beta(-\omega_n)/\d\omega_n$,
whereas, as usual, $A'_\beta(\omega_n) = \d A_\beta(\omega_n)/\d\omega_n$ etc.]
Therefore, we define
\begin{align}
  \Sigma(\omega_n) &= A_\beta(\omega_n) + A_\beta(-\omega_n)
  \nonumber \\
  &= 2\IntW\Jw\frac{\omega}{\omega^2-\omega_n^2} \\
  \Delta_\beta(\omega_n) &= A_\beta(\omega_n) - A_\beta(-\omega_n)
  \nonumber \\
  &= 2\omega_n\IntW\Jw\frac{1}{\omega^2-\omega_n^2}
    \coth(\tfrac{1}{2}\beta\hbar\omega) \\
  \Delta'_\beta(\omega_n) &= A'_\beta(\omega_n) + A'_\beta(-\omega_n)
  \nonumber \\
  &= 2\IntW\frac{\Jw}{\hbar}\frac{(\omega^2+\omega_n^2)}{(\omega^2-\omega^2_n)^2}
    \coth(\tfrac{1}{2}\beta\hbar\omega) \\
  \Sigma'(\omega_n) &= A'_\beta(\omega_n) - A'_\beta(-\omega_n)
  \nonumber \\
  &= 4\omega_n\IntW\frac{\Jw}{\hbar}
  \frac{\omega}{(\omega^2-\omega_n^2)^2}.
\end{align}


\subsubsection{Normalising the second order \MFGS}

From~\eqref{eq:sm:8} we get the second order partition function
\begin{equation}\label{eq:sm:15}
  \ZSMFwk = \tr[\rhowkun]
    = 1 + \beta\lambda^2\sum_{n}\tr\left[\tauS\Xop_n\Xop_n^\dagger\right]
      A_\beta(\omega_n).
\end{equation}
This can be used directly to evaluate the second order expectation value
$\expval{\Sz}^{(2)}$, but instead we will proceed to derive the second order
\MFGS.
This normalised state can be arrived at
in two ways. First we can write
$\rhowk = \rhowkun/\ZSMFwk$,
and then, on the basis that the second order correction
in~\eqref{eq:sm:15} is $\ll 1$, we can make the
binomial approximation
\begin{equation}
  \frac{1}{\ZSMFwk} = 1 -
    \beta\lambda^2\sum_{n}\tr\left[\tauS\Xop_n\Xop_n^\dagger\right]
    A_\beta(\omega_n),
\end{equation}
in which case the normalised state is, correct to second order
\begin{align}
  \rhowk
  &= \tauS
  \nonumber \\
  &+ \lambda^2\sum_{n}\Bigg[\left[\Xop_n^\dagger,\tauS\Xop_n\right]
    A'_\beta(\omega_n)
  \nonumber \\
  &\quad\quad+ \beta\tauS\left(\Xop_n\Xop_n^\dagger -
    \trS\left[\tauS\Xop_n\Xop_n^\dagger\right]\right)A_\beta(\omega_n)\Bigg]
  \nonumber \\
  &+ \lambda^2\sum_{m \ne n}
    \left(\left[\Xop_m,\Xop_n^\dagger \tauS\right] +
    \left[\tauS\Xop_n,\Xop_m^\dagger\right]\right)\frac{A_\beta(\omega_n)}{\omega_{nm}}.
  \label{eq:sm:NormalizedMFGS}
\end{align}
The issue with this approach is that it assumes the validity of the
binomial approximation, which requires the second order correction to 
$\ZSMFwk$ to be $\ll 1$. However, this term grows linearly with
$\beta$, so that at a sufficiently low temperature this second
order correction will exceed unity by an arbitrary amount, and the
binomial approximation cannot be justified.

An alternate approach is to deal directly with the exact density operator
\begin{equation}
  \tsysMFGS(\lambda) = \frac{\rhoun(\lambda)}{\ZSMFqu(\lambda)},
\end{equation}
where the dependence on $\lambda$ is made explicit, and write
\begin{equation}
  \tsysMFGS(\lambda) = \tsysMFGS(0) +
    \tfrac{1}{2}\lambda^2\frac{\d^2 \tsysMFGS}{\d\lambda^2}(0) +
    \order(\lambda^4),
\end{equation}
where $\tsysMFGS(0) = \tauS$ is the Gibbs state of the system in the limit of
vanishingly small system-reservoir coupling, and it has been recognised that odd
order contributions will vanish.

From this we also find
\begin{equation}
  \ZSMFqu(\lambda) = 1 +
    \tfrac{1}{2}\lambda^2\frac{\d^2\ZSMFqu}{\d\lambda^2}(0) +
    \order(\lambda^4).
\end{equation}
If we now do a Taylor series expansion of $\tsysMFGS(\lambda)$ we find, using
$\ZSMFqu(0) = 1$,
\begin{align}
  \tsysMFGS(\lambda)
  &= \tauS + \tfrac{1}{2}\lambda^2\left(\frac{\d^2\rhoun}{\d\lambda^2}(0) -
    \frac{\d^2\ZSMFqu}{\d\lambda^2}(0)\tauS\right)
  \nonumber \\
  &+ \order(\lambda^4)
  \nonumber \\
  &= \tauS + \lambda^2\sum_{n}\Bigg[\left[\Xop_n^\dagger,\tauS\Xop_n\right]
    A'_\beta(\omega_n)
  \nonumber \\
  &\quad\quad + \beta\tauS\left(\Xop_n\Xop_n^\dagger -
    \tr\left[\tauS\Xop_n\Xop_n^\dagger\right]\right)A_\beta(\omega_n)\Bigg]
  \nonumber \\
  &+ \sum_{m \ne n} \left(\left[\Xop_m,\Xop_n^\dagger
    \tauS\right] +
    \left[\tauS\Xop_n,\Xop_m^\dagger\right]\right)\frac{A_\beta(\omega_n)}{\omega_{nm}}
  \nonumber \\
  &+ \order(\lambda^4).
\end{align}
So we regain~\eqref{eq:sm:NormalizedMFGS}, but without having to consider any
restrictions on $\beta$. In contrast the binomial expansion based derivation
seems to imply that irrespective of the choice of coupling strength, there will
always be a temperature below which the binomial approximation will fail
and~\eqref{eq:sm:NormalizedMFGS} can lead to incorrect results below this
temperature. But this argument cannot be sustained as the validity of the second
order expansion is not constrained by any lower temperature limit implied by the
binomial expansion as it can be obtained without making this approximation.

\smallskip

What we now have is the necessary requirement that (for some definition of the
norm $||\ldots||$ of the operators involved)
\begin{equation} \label{eq:sm:24}
  \tfrac{1}{2}\lambda^2\left|\left|\left( \frac{\d^2\tsysMFGS}{\d\lambda^2}(0)
  - \tauS\frac{\d^2\ZSMFqu}{\d\lambda^2}(0)\right)\right|\right|
  \ll \left|\left|\tauS\right|\right|,
\end{equation}
for the second order result~\eqref{eq:sm:NormalizedMFGS} to be valid. This of
course is not a sufficient condition as the higher order terms,
$\order(\lambda^4)$, are not guaranteed to be negligible.

The concern is the low temperature limit $\beta\to\infty$, where the term linear
in $\beta$ seems to imply linear divergence so the condition~\eqref{eq:sm:24}
cannot be met. However, it can be shown that in this limit the second order
correction term in~\eqref{eq:sm:NormalizedMFGS} actually
vanishes~\cite{cresser2021a}. It also does so for $\beta \to 0$, the high
temperature limit, so there might be an intermediate temperature for which the
condition~\eqref{eq:sm:24} is not satisfied, this then requiring a weaker
interaction coupling strength.

The conclusion then is that for sufficiently weak coupling, the
result~\eqref{eq:sm:NormalizedMFGS} will hold true for all temperatures.



\bigskip

To evaluate the second order expression for the normalised density operator
given by~\eqref{eq:sm:NormalizedMFGS} we need to expand $\Xop=S_\theta$ in terms
of the energy eigenoperators $\Xop_n$,
\begin{align}
  \Xop &= \Sz\cos\theta - \Sx\sin\theta \\ \nonumber
    &= -\tfrac{1}{2}\sin\theta\Sm +\cos\theta\Sz - \tfrac{1}{2}\sin\theta\Sp,
\end{align}
so we can identify, from $\Xop = \Xop_{-1} + \Xop_0 + \Xop_{+1}$,
\begin{align}
  \Xop_{-1} &= -\tfrac{1}{2}\sin\theta\Sm, \\
  \Xop_0 &= \cos\theta\Sz, \\
  \Xop_{+1} &= -\tfrac{1}{2}\sin\theta\Sp.
\end{align}
To determine the corresponding eigenfrequencies, we use
$\left[\HS,\Xop_n\right] = \omega_n\Xop_n$ and find that
\begin{equation}
  \left[\HS,\Xop_{-1}\right] =
  \left[-\wL\Sz,-\tfrac{1}{2}\sin\theta\Sm\right] = \wL\Xop_{-1},
\end{equation}
and hence $\omega_{-1} = \wL$. It follows that $\omega_{+1} = -\wL$, and by
inspection, $\omega_0 = 0$.

To evaluate $\rhowk$ we then have a number of sums to evaluate, and from
that expression we can then calculate the expectation values of $\Sz$ and 
$\Sx$. The calculation of these quantities is made `easier' by the fact that
$\tauS$ is diagonal in the $\Sz$ basis, and that $\expval{\Sy} = 0$.
After somewhat lengthy but straightforward calculations we find that
\begin{align}
  \expval{\Sz}^{(2)}
  &= \expval{\Sz}_0
  + \tfrac{1}{4}\hbar\lambda^2\sin^2\theta\Big[(\SO(\SO+\hbar)
  \nonumber \\
  &\qquad - \expval{\Sz^2}_0) \Sigma'(\wL) - \expval{\Sz}_0\hbar\Delta'_\beta(\wL)\Big]
  \nonumber \\
  &- \beta\lambda^2\Big[\tfrac{1}{4}\sin^2\theta
    \big(\left(\expval{\Sz^2}_0 - \expval{\Sz}^2_0\right)\hbar\Delta_\beta(\wL)
  \nonumber \\
  &\qquad + \left(\expval{\Sz^3}_0 - \expval{\Sz}_0\expval{\Sz^2}_0\right)\Sigma(\wL)\big)
  \nonumber \\
  &\qquad - \cos^2\theta \left(\expval{\Sz^3}_0 - \expval{\Sz}_0\expval{\Sz^2}_0\right)\Q\Big],
  \label{eq:sm:SzQuantum}
\end{align}
and
\begin{align}
  \expval{\Sx}^{(2)}
  &= \lambda^2\frac{\sin2\theta}{4\wL}
    \Big[\left(\SO(\SO+\hbar) -
    \expval{\Sz^2}_0\right)\Sigma(\wL)
  \nonumber \\
  &\qquad - \hbar\expval{\Sz}_0\Delta_\beta(\wL) -
    4\expval{\Sz^2}_0\Q\Big],
  \label{eq:sm:SxQuantum}
\end{align}
where $\expval{\ldots}_0 = \tr[\tauS\ldots]$.


\subsection{Quantum to classical limit for weak coupling}

In what follows we explicitly verify the quantum to classical transition in the
large spin limit presented in~\ref{sec:quantumtoclassical}, using the quantum
and classical weak coupling expressions found in the previous sections.

First,~\eqref{eq:sm:SzQuantum}, we have $\expval{\Sz}$, regrouped to read 
\begin{align}
    \expval{\Sz}
    &= \langle S_z\rangle_0+\tfrac{1}{4}\lambda^2\sin^2\theta\Big[
	    (S_0(S_0+\hbar)-\langle S_z^2\rangle_0)\hbar\Sigma' 
    \nonumber\\
	&-\langle S_z\rangle_0\hbar^2\Delta'_\beta 
	-\beta\big(\left(\langle S_z^2\rangle_0
	-\langle S_z\rangle^2_0\right)\hbar\Delta_\beta
	\nonumber \\
	&+\left(\langle S_z^3\rangle_0-\langle S_z\rangle_0\langle S_z^2\rangle_0\right)(\Sigma-2Q)\big)\Big]
	\nonumber\\
	&+\tfrac{1}{4}\beta\lambda^2(1+3\cos2\theta) Q\left(\langle S_z^3\rangle_0-\langle S_z\rangle_0\langle S_z^2\rangle_0\right).
\end{align}
Introducing the scaled temperature $\beta' = \beta\SO$ and the scaled spin
$s_z=S_z/S_0$ and taking the limit $\SO\to\infty$ with $\beta'$ held constant
gives
\begin{align}
    \langle s_z\rangle
    &= \langle s_z\rangle_0+\tfrac{1}{4}\lambda^2\sin^2\theta
	\Big[(1-\langle s_z^2\rangle_0)\hbar(S_0\Sigma') 
	\nonumber \\
	&-\langle s_z\rangle_0\hbar^2\Delta'_\beta 
	    -\beta'\big(\left(\langle s_z^2\rangle_0
	    -\langle s_z\rangle^2_0\right)\hbar\Delta_\beta
	\nonumber \\
	&+\left(\langle s_z^3\rangle_0-\langle s_z\rangle_0\langle s_z^2\rangle_0\right)((S_0\Sigma)-2S_0Q)\big)\Big]
	\nonumber\\
    &+\tfrac{1}{4}\beta'\lambda^2(1+3\cos2\theta)(S_0 Q)\left(\langle s_z^3\rangle_0-\langle s_z\rangle_0\langle s_z^2\rangle_0\right).
\end{align}
with (and noting that $\SO\Jw$ is independent of $\SO$)
\begin{align}
    \SO\Sigma&\to \int_{0}^{\infty}(\SO\Jw)
        \frac{2\omega}{\omega^2-\wL^2}\mathrm{d}\omega, \\
    \Delta_\beta&\to \int_{0}^{\infty}\frac{(\SO\Jw)}{\hbar}
        \frac{4\wL}{\omega^2-\wL^2}\frac{1}{\beta'\omega}\mathrm{d}\omega, \\
    \Delta'_\beta&\to \int_{0}^{\infty}\frac{(\SO\Jw)}{\hbar^2}
        \frac{4\left(\omega^2 + \wL^2\right)}{\left(\omega^2-\wL^2\right)^2}
        \frac{1}{\beta'\omega}\mathrm{d}\omega, \\
    \SO\Sigma'&\to \int_{0}^{\infty}\frac{(\SO\Jw)}{\hbar}
        \frac{4\wL\omega}{\left(\omega^2-\wL^2\right)^2}\mathrm{d}\omega, \\
    \SO Q&\to \int_{0}^{\infty}(\SO\Jw)\frac{1}{\omega}\mathrm{d}\omega.
\end{align}
Making use of the $S_0\to\infty$ limit of $\langle s_z^n\rangle_0$, $n=1,2,3$
with $\beta'$ held constant, given from
\eqref{eq:sm:ZerothOrderQuantumExpectationValues} by the classical forms
\eqref{eq:sm:ZerothOrder}:
\begin{align}
    \langle s_z\rangle_0 &=
        \coth(\beta' \omega_\text{L})-\frac{1}{\beta' \omega _\text{L}}, \\
    \langle s_z^2\rangle_0 &=
        1-\frac{2\coth(\beta' \omega_\text{L})}{\beta' \omega_\text{L}}+\frac{2}{(\beta' \omega_\text{L})^2}, \\
    \langle s_z^3\rangle_0 &=
        \coth(\beta'\omega_\text{L})-\frac{3}{\beta' \omega_\text{L}}+\frac{6\coth(\beta' \omega_\text{L})}{(\beta' \omega_\text{L})^2} \nonumber \\
        &-\frac{6}{(\beta' \omega_\text{L})^3},
\end{align}
and the above limiting forms for the integrals, we find
that the factor multiplying $\sin^2\theta$ vanishes and we are left with
\begin{align}
  \langle s_z\rangle &=
  \langle s_z\rangle_0 \\ \nonumber
  &+\frac{1}{4}\beta'\lambda^2
  S_0Q(1+3\cos2\theta)
  \left(\langle s_z^3\rangle_0-\langle s_z\rangle_0\langle s_z^2\rangle_0\right),
\end{align}
which  on substituting for the $\langle s_z^n\rangle_0$ yields a
result formally identical to the fully classical 
result,~\eqref{eq:sm:SzClassical1}. In a similar way we can check
the large spin limit for $\langle S_x\rangle$, and we regain the
classical result,~\eqref{eq:sm:SxClassical2}.


\section{Ultrastrong coupling limit}
\label{sec:ultrastrong}

In this section the we examine ultrastrong coupling limit of the quantum and
classical \MFGS.


\subsection{Classical ultrastrong coupling limit}
\label{sm:CMFUS}

The ultrastrong limit is the limit in which the coupling $\lambda$
is made very large, in principle taken to infinity.
To take this limit, first note that the partition function can be written as
\begin{equation}
  \Z_\text{S}^\text{cl} = \ZO^\text{cl}\int_{0}^{\pi}\mathrm{d}\theta'\sin\theta'     
      e^{\beta\wL\SO\cos\theta'}\,F(\lambda,\theta,\theta'),
\end{equation}
where
\begin{align}
  &F(\lambda,\theta,\theta') = \\ \nonumber
  &\int_{0}^{2\pi}d\varphi'
  e^{\tfrac{1}{2}\beta\lambda^2\SO^2Q
     \left(\sin\theta\sin\theta'\cos\varphi'-\cos\theta\cos\theta'\right)^2}.
\end{align}


Defining $a = \tfrac{1}{2}\beta Q\SO^2$, expanding the exponent and using the
periodicity of the trigonometric functions we can rewrite
\begin{align}
  &F(\lambda,\theta,\theta')
  \nonumber \\
  &=e^{a\lambda^2\cos^2(\theta'-\theta)}
      H(\cos\theta'\cos\theta)
      \int_{0}^{2\pi}\mathrm{d}\varphi'
  \nonumber \\
  &\quad e^{-4a\lambda^2\sin\theta\sin\theta'\cos^2(\varphi'/2)
      \big(\sin\theta\sin\theta'\sin^2(\varphi'/2)}
  \nonumber \\
  &\qquad\qquad\qquad\qquad\qquad\qquad {}^{+ \cos\theta\cos\theta'\big)}
  \nonumber \\
  &+ e^{a\lambda^2\cos^2(\theta'+\theta)}
      H(-\cos\theta'\cos\theta)
      \int_{-\pi}^{\pi}\mathrm{d}\varphi'
  \nonumber \\
  &\quad e^{-4a\lambda^2\sin\theta\sin\theta'\sin^2(\varphi'/2)
      \big(\sin\theta\sin\theta'\cos^2(\varphi'/2)}
  \nonumber \\
  &\qquad\qquad\qquad\qquad\qquad\qquad{}^{-\cos\theta\cos\theta'\big)}
\end{align}
where $H(x)$ is the Heaviside step function.
The advantage of this rewriting of $F(\lambda,\theta,\theta')$ is that now
the exponents in the integrands are all negative (or zero) over the range
of integration.
The exponent of the integrand for the first integral where $\cos\theta'\cos\theta > 0$
will vanish at $\varphi' = \pi$, while for the second integral,
where $\cos\theta'\cos\theta < 0$, the exponent of the integrand will vanish at
$\varphi' = 0,2\pi$. At these points the integrands will have local maxima
which will become increasingly sharp as $\lambda$ is increased. 
Similarly, for the second integral the maximum of the second integrand lies at
$\varphi' = 0$.

Thus, as $\lambda$ is increased, we can approximate the exponent in the integral
by its behaviour in the neighbourhood of $\varphi' = \pi$ for the first integral,
and in the neighbourhood of $\varphi' = 0$ for the second one.
This is just using the method of steepest descent.
We then obtain
\begin{align}
  \Z_\text{S}^\text{cl} &\sim \ZO^\text{cl} e^{a\lambda^2}\int_{0}^{\pi}d\theta'\sin\theta'
      e^{\beta \omega_\text{L}S_0\cos\theta'}
  \Big[
  \nonumber \\
  &e^{-a\lambda^2\sin^2(\theta'-\theta)}
     R_-(\theta',\theta)
      H(\cos\theta'\cos\theta)
  \nonumber \\
  &\qquad\qquad\int_{0}^{2\pi}\mathrm{d}\varphi'\delta(\varphi'-\pi)
  \nonumber \\
  &+ e^{-a\lambda^2\sin^2(\theta'+\theta)}
     R_+(\theta',\theta)
      H(-\cos\theta'\cos\theta)
  \nonumber \\
  &\qquad\qquad\int_{-\pi}^{\pi}\mathrm{d}\varphi'\delta(\varphi')\Big].
\end{align}
where
\begin{equation}
    R_\pm(\theta',\theta)= \sqrt{\frac{\pi}{a\lambda^2|\sin\theta'\sin\theta\cos(\theta'\pm\theta)|}},
\end{equation}
and for later interpretation purposes, the $\varphi'$ integrals have been
retained unevaluated.

Once again we notice that the exponents in the integrands are all negative.
The zeroes of the exponents will occur within the range of integration for
$\theta' = \theta$ for the first exponents, and for $\theta' = \pi - \theta$
for the second.
Therefore, in the large $\lambda$ limit we have
\begin{align}
  \ZScl\to\ZCMFUS &\sim
  \ZO^\text{cl}\frac{\pi e^{a\lambda^2}}{a\lambda^2}
  \int_{0}^{\pi}\mathrm{d}\theta' e^{\beta\wL\SO\cos\theta'}
  \Big[
  \nonumber \\
  &\int_{0}^{2\pi}\mathrm{d}\varphi' \delta(\theta'-\theta)\delta(\varphi'-\pi)
  \nonumber \\
  &+ \int_{-\pi}^{\pi}\mathrm{d}\varphi'\delta(\theta'+\theta-\pi)\delta(\varphi')\Big].
\end{align}
This suggests that in the large $\lambda$ limit, the spin orients
itself in either the $\theta' = \theta, \varphi' = \pi$ or
$\theta' = \pi-\theta, \varphi' = 0$ directions, though with different
weightings for the two directions.

If we return to the interaction on which this result is based, that is
\begin{align}
  V &= (\Sz\cos\theta-\Sx\sin\theta)\Bop \\ \nonumber
    &= \mathbf{S}\cdot\Bop(-\sin\theta\mathbf{x} + \cos\theta\mathbf{z})
     = \mathbf{S}\cdot\mathbf{\Bop},
\end{align}
we see that the vector $-\sin\theta\mathbf{x} + \cos\theta\mathbf{z}$ has the polar
angles $\theta' = \theta,\varphi' = \pi$. But as $\Bop$ can be fluctuate between
positive or negative values, the vector $\mathbf{\Bop}$ can fluctuate between this
and the opposite direction $\theta' = \pi-\theta, \varphi' = 0$.
So the effect of the ultrastrong noise is to force the spin to orient itself in
either of these two directions.


%
Returning to the expression for the partition function, we have
\begin{align}\label{eq:sm:ZCMFUS}
  \ZCMFUS &\sim \ZO''\left(e^{\beta\wL\SO\cos\theta} + 
    e^{-\beta\wL\SO\cos\theta}\right) \nonumber \\
  &\propto\cosh(\beta\wL\SO\cos\theta),
\end{align}
where extraneous factors have been absorbed into $\ZO''$.
Further, to get at $\expval{\Sz}$ we need
\begin{align}
  \expval{\Sz}
  &= \SO\frac{\int_{0}^{\pi}\mathrm{d}\theta'\cos\theta'
      F(\lambda,\theta',\theta)}{\int_{0}^{\pi}\mathrm{d}\theta' 
      F(\lambda,\theta',\theta)} \nonumber \\
  &= \SO\cos\theta\tanh(\beta\wL\SO\cos\theta).
\end{align}
These results are of the same form as found for a quantum spin half.
That result is understandable given that the spin half would have two
orientations, which mirrors the two orientations that emerge in the strong
coupling limit here in the classical case.


\subsection{Quantum ultrastrong coupling limit}
\label{sm:QMFUS}


The aim here is to derive an expression for the quantum MFG state of a spin
$\SO$ particle coupled to a thermal reservoir at a temperature $\beta^{-1}$,
\eqref{eq:sm:QMFGS}.

The ultrastrong coupling limit is achieved by making $\lambda$ very much
greater than all other energy parameters of the system, in effect,
$\lambda\to\infty$. However, note the absence of the `counter-term'
$-\lambda^2(\cos\theta\Sz-\sin\theta\Sx)^2Q$ in the above Hamiltonian. This
term appears in \cite{cresser2021a}, where it is found to be cancelled in the
strong coupling limit when the trace over the reservoir states is made. Here,
that cancellation will not take place, so its presence must be taken into account.
It will have no impact in the case of $\SO = \tfrac{1}{2}$, as this will be a
c-number contribution, but it will have an impact otherwise.

With $\HS = -\wL\Sz$ and $P_{s_\theta} = |\stheta\rangle\langle\stheta|$ the
projector onto the eigenstate $|\stheta\rangle$ of $\Stheta$ where
\begin{equation}
  \Stheta = \cos\theta\Sz - \sin\theta\Sx;
  \quad \Stheta|\stheta\rangle = \stheta|\stheta\rangle
\end{equation}
we have, in the ultrastrong coupling limit, the unnormalised MFG state of the
particle 
\begin{align}
  \tilde{\rho}
  &= \exp\left[-\beta\sum_{\stheta = -\SO}^{\SO} P_{\stheta} \HS P_{\stheta}\right]
      e^{\beta\lambda^2\Stheta^2Q}
  \nonumber \\
  &= \sum_{\stheta = -\SO}^{\SO} P_{\stheta} \exp\left[-\beta\langle\stheta| \HS |\stheta\rangle\right]
      e^{\beta \lambda^2\gamma^2s_\theta^2Q}.
\end{align}
Note, as a consequence of the absence of a counter-term, the contribution
$\exp[\beta\lambda^2\Stheta^2Q]$ is not cancelled.

Further note the limits on the sum are $\pm\SO$. This follows since $\Stheta =
\cos\theta\Sz - \sin\theta\Sx$ is just $\Sz$ rotated around the $y$ axis, i.e.,
\begin{equation}
  \cos\theta\Sz - \sin\theta\Sx = e^{i\theta\Sy}\Sz e^{-i\theta\Sy} = \Stheta
\end{equation}
so the eigenvalue spectrum of $\Stheta$ will be the same as that of $\Sz$,
i.e., $\sz = -\SO,-\SO+1,\ldots,\SO-1,\SO$. The eigenvectors of $\Stheta$ are
then, from $\Sz|\sz\rangle = \sz|\sz\rangle$ 
\begin{equation}
  \Stheta e^{i\theta\Sy}|\sz\rangle = \sz\,e^{i\theta\Sy}|\sz\rangle
\end{equation}
i.e., the eigenvectors of $\Stheta$ are
\begin{equation}
  |\stheta\rangle = e^{i\theta\Sy}|\sz\rangle;
  \quad \stheta = \sz = -\SO,\ldots,\SO.
\end{equation}
We then have
\begin{align}
  \langle\stheta|\HS|\stheta\rangle
      &= \wL\langle\sz|e^{-i\theta\Sy}\Sz e^{i\theta\Sy}|\sz\rangle
      \nonumber \\
      &= \wL\langle\sz|\cos\theta\Sz + \sin\theta\Sx|\sz\rangle
      \nonumber \\
      &= \wL\sz\cos\theta,
\end{align}
from which follows
\begin{equation}
  \tilde{\rho}
  = \sum_{\sz = -\SO}^{\SO}e^{i\theta\Sy}|\sz\rangle\langle \sz|e^{-i\theta\Sy}
      e^{\beta\wL\sz\cos\theta}e^{\beta\lambda^2Q\sz^2}.
\end{equation}
The partition function is then given by
\begin{equation}
  \tilde{\Z}_\text{S}^\text{qu} = \sum_{\sz = -\SO}^{\sz = \SO}e^{\beta\wL\sz\cos\theta}
      e^{\beta\lambda^2Q\sz^2}.
\end{equation}
This cannot be evaluated exactly, but the limit of large $\lambda$ is yet to be
taken. The dominant contribution to the sum in that limit will be for $\sz =
\pm\SO$, so we can write
\begin{align}
  \ZQMFUS e^{-\beta\lambda^2Q\SO^2}
  &\sim e^{\beta\wL\SO\cos\theta} + e^{-\beta\wL\SO\cos\theta}
  \nonumber \\
  &\propto\cosh(\beta\wL\SO\cos\theta).
\end{align}
Apart from an unimportant proportionality factor, this is exactly the same results as found for the classical case in the limit of ultrastrong coupling, \eqref{eq:sm:ZCMFUS}.

In fact, if we follow the same procedure as above for the normalised density
operator, we have
\begin{equation}
  \rho = \frac{\sum_{\sz = -\SO}^{\SO}e^{i\theta\Sy}|\sz\rangle\langle\sz|e^{-i\theta\Sy}
      e^{\beta\wL\sz\cos\theta} e^{\beta\lambda^2Q\sz^2}}
      {\sum_{\sz = -\SO}^{\sz = \SO} e^{\beta\wL\sz\cos\theta}
      e^{\beta\lambda^2Q\sz^2}}.
\end{equation}
The dominant contribution in the limit of large $\lambda$ will then be for
$\sz = \pm\SO$, so we get
\begin{align}
  \rho &= \frac{e^{i\theta\Sy}|\SO\rangle\langle\SO|e^{-i\theta\Sy}e^{\beta\wL\SO\cos\theta}}{e^{\beta\wL\SO\cos\theta} + e^{-\beta\wL\SO\cos\theta}}
  \nonumber \\
  &+ \frac{e^{i\theta\Sy}|{-\SO}\rangle\langle -\SO|e^{-i\theta\Sy}e^{-\beta\wL\SO\cos\theta}}{e^{\beta\wL\SO\cos\theta} + e^{-\beta\wL\SO\cos\theta}}.
\end{align}
To work this out for arbitrary spin would require introducing the matrix
elements of the rotation matrices for $\exp[-i\theta\Sy]$, which would be a
complex business. Instead we will simply introduce
\begin{equation}
  \sigma_\pm(\theta)
  = e^{i\theta\Sy} 
      \left(|\SO\rangle\langle\SO|\pm|{-\SO}\rangle\langle -\SO|\right)
    e^{-i\theta\Sy},
\end{equation}
and write this as
\begin{equation}
  \rho = \tfrac{1}{2}\left[\sigma_+(\theta) + \sigma_-(\theta)\tanh(\beta\lambda\wL\SO\cos\theta)\right].
\end{equation}
For $\SO = \tfrac{1}{2}$ this above general result reduces to the earlier
obtained result (for which the counter-term can be safely ignored), which is
\begin{equation}
  \rho = \tfrac{1}{2}\left[1 + \sigma(\theta)\tanh(\tfrac{1}{2}\beta\wL\cos\theta)\right],
\end{equation}
where $\sigma_+(\theta) = 1$ and $\sigma(\theta) = \cos\theta\sigma_z -
\sin\theta\sigma_x$ and for which $\expval{\Sz}
=\tfrac{1}{2}\cos\theta\tanh(\tfrac{1}{2}\beta\lambda\wL\cos\theta)$ and
$\expval{\Sx} =
\tfrac{1}{2}\sin\theta\tanh(\tfrac{1}{2}\beta\lambda\wL\cos\theta)$.



\end{document}